\documentclass[aps,prb,amsmath,amssymb,amsfonts,floatfix,superscriptaddress,twocolumn,nofootinbib]{revtex4-1}                                                                                    
\usepackage{graphicx}
\usepackage{color}
\usepackage[utf8]{inputenc}
\usepackage{dcolumn}
\usepackage{hyperref}
\usepackage{textcomp}
\usepackage{enumitem}
\usepackage{color}
\usepackage{colortbl}
\usepackage{xcolor}
\usepackage{listings}
\usepackage{braket}

\definecolor{linkcolor}{HTML}{799B03}
\definecolor{urlcolor}{HTML}{799B03}
\hypersetup{pdfstartview=FitH, linkcolor=linkcolor,urlcolor=urlcolor,colorlinks=true}

\hypersetup{
    unicode=false,          
    pdftoolbar=true,        
    pdfmenubar=true,        
    pdffitwindow=false,     
    pdfstartview={FitH},    
    pdfauthor={},     		
    colorlinks=true,       	
    linkcolor=blue,         
    citecolor=red,        	
}

\usepackage{amstext}
\usepackage{placeins}
\usepackage{epstopdf}
\usepackage{multirow}
\usepackage{tabularx}
\usepackage{tikz,bm}
\usetikzlibrary{positioning,arrows}

\graphicspath{{figures/}}

\newcommand{\kv}{\ensuremath{\mathbf{k}}}
\newcommand{\qv}{\ensuremath{\mathbf{q}}}

\usetikzlibrary{arrows,shapes.geometric,backgrounds,positioning,calc}

\usepackage{todonotes}

\begin{document}

\title{Exact diagonalization solver for the extended dynamical mean-field theory}

\author{Darya Medvedeva}
\thanks{These authors contributed equally}
\affiliation{Theoretical Physics and Applied Mathematics Department, Ural Federal University, Mira Str.19, 620002 Ekaterinburg, Russia}
\author{Sergei Iskakov}
\thanks{These authors contributed equally}
\affiliation{Theoretical Physics and Applied Mathematics Department, Ural Federal University, Mira Str.19, 620002 Ekaterinburg, Russia}
\affiliation{Department of Physics, University of Michigan, Ann Arbor, Michigan 48109, USA}
\author{Friedrich Krien}
\thanks{These authors contributed equally}
\affiliation{Institute for Theoretical Physics, University of Hamburg, Jungiusstrasse 9, 20355 Hamburg, Germany}
\author{Vladimir V. Mazurenko}
\affiliation{Theoretical Physics and Applied Mathematics Department, Ural Federal University, Mira Str.19, 620002 Ekaterinburg, Russia}
\author{Alexander I. Lichtenstein}
\affiliation{Theoretical Physics and Applied Mathematics Department, Ural Federal University, Mira Str.19, 620002 Ekaterinburg, Russia}
\affiliation{Institute for Theoretical Physics, University of Hamburg, Jungiusstrasse 9, 20355 Hamburg, Germany}
\date{\today}

\begin{abstract}
We present an efficient exact diagonalization scheme for the extended dynamical mean-field theory and
apply it to the extended Hubbard model on the square lattice with nonlocal charge-charge interactions.
Our solver reproduces the phase diagram of this approximation with good accuracy.
Details on the numerical treatment of the large Hilbert space of the auxiliary Holstein-Anderson impurity problem are provided.
Benchmarks with a numerically exact strong-coupling continuous-time quantum-Monte Carlo solver
show better convergence behavior of the exact diagonalization in the deep insulator.
Special attention is given to possible effects due to the discretization of the bosonic bath.
We discuss the quality of real axis spectra and address the question of screening in the Mott insulator within extended dynamical mean-field theory.
\end{abstract}

\maketitle


The description of strong correlations is a challenging topic in the fields of spintronics, nanoelectronics, and molecular electronics.
For future applications some typical effects of strong correlations are potentially attractive, such as the Mott metal-insulator
transition,~\cite{Mott_MeIns_Trans_book, RevModPhys.70.1039,PhysRevB.89.165138,PhysRevB.94.205110, PhysRevB.90.195114} the Kondo
resonance,~\cite{Allen_Kondo_resonance} high-temperature superconductivity,~\cite{RevModPhys.66.763} and itinerant ferromagnetism.~\cite{PhysRevB.58.12749}
Mott physics has been suggested for applications in the context of field effect transistors~\cite{ doi:10.1080/10408436.2012.719131}
and memory devices.~\cite{10.1155/2014/927696}

Approaches based on the dynamical mean-field theory (DMFT)~\cite{GeorgesKotliar, PhysRevLett.62.324} have been very successful in the description of strong correlations.
In case of sufficiently short-ranged interactions, strongly correlated systems are often described by an effective Hubbard model. 
In single-site DMFT one approximates the Hubbard model by solving a local impurity problem instead, where the impurity can be seen as a representative of each single atom of the lattice.
The hopping of electrons from the impurity to neighboring atoms and vice versa is modeled by a self-consistent hybridization of the impurity with an effective bath.
For the Hubbard model this description is suitable because the interaction, as well as the dominant correlations, are local.
Methods based on DMFT have been used to predict strong correlation effects in real materials.~\cite{RevModPhys.78.865, Vollhardt2017}

Exact diagonalization (ED) has been used early on as a solver for the local reference system of DMFT, the Anderson impurity
model.~\cite{Lin_Gubernatis, PhysRevLett.72.1545, S0217984994000571, GeorgesKotliar, PhysRevB.86.165128, PhysRevB.88.035123} ED methods
operate with a finite Hamiltonian, whose size is a limiting factor to their applicability.
As a consequence, the effective hybridization function needs to be "discretized", i.e., the number of energy levels of the bath needs to be finite and small, while it is infinite in the thermodynamic limit.
However, few fermionic bath levels are indeed needed to describe the Hubbard model with good precision.~\cite{PhysRevLett.72.1545}
The noise-free solution of the Anderson impurity model on the real axis and the absence of any sign problem make ED solvers an alternative
in circumstances where Continuous-Time Quantum Monte Carlo (CTQMC) solvers~\cite{Werner} prove unsuitable.

An important focus in the field of strongly correlated materials is taking long-ranged interactions into
account.~\cite{0953.8984.28.38.383001, Kristoffel2017}
Some adatom systems~\cite{PhysRevB.94.214411, PhysRevLett.110.166401, PhysRevB.94.224418} show very strong nonlocal electrostatic and magnetic interactions, 
due to a spatial delocalization of the Wannier functions at the Fermi level.
In some cases the estimated value of the Coulomb repulsion between neighboring atoms is only two times smaller than
on-site.~\cite{PhysRevB.94.214411} Sometimes an optimal mapping of a system with a long-ranged interactions to an effective Hubbard model
is possible.~\cite{PhysRevLett.111.036601}
However, the repulsive Hubbard model cannot describe essential effects of nonlocal interactions like the charge-order.

There are several ways to take non-local interactions and correlations into account.
A commonly used method is to consider a cluster of atoms, instead of a single atom, as the smallest unit of the crystal.
This idea underlies a number of self-consistent cluster approaches, e.g., the cellular DMFT
(CDMFT),~\cite{PhysRevB.62.R9283,PhysRevLett.87.186401} the dynamical cluster approximation
(DCA),~\cite{PhysRevB.58.R7475,PhysRevB.95.115149} and the variational cluster approximation.~\cite{Potthoff_VDA, Potthoff_VDA2,
PhysRevLett.91.206402}
However, the calculation of the correlation functions of a hybridized cluster, such as in CDMFT, requires substantially more computational
effort than the solution of a single-impurity problem and costs further increase with the cluster size.
Furthermore, the range of non-local interactions and correlations is restricted to the cluster size.

Another option to take nonlocal interactions into account is the extended DMFT (EDMFT).~\cite{PhysRevLett.77.3391, PhysRevB.61.5184,
PhysRevLett.84.3678, PhysRevB.63.115110,PhysRevB.66.045111}
In this approach one maintains the single-impurity framework of DMFT by augmenting its local reference system by a retarded interaction.
This nonlocal-in-time interaction is fixed by a two-particle self-consistency condition, in analogy to the self-consistent determination of the hybridization function in DMFT.
A number of methods for the solution of the resulting Holstein-Anderson impurity model exist, e.g., the Numerical Renormalization
Group,~\cite{0953-8984-10-37-021, PhysRevB.68.045106, PhysRevB.85.075113} Discrete Time QMC~\cite{PhysRevLett.56.2521} or
CTQMC.~\cite{Werner, Otsuki}
A precondition for the use of EDMFT is that even in the context of nonlocal interactions, the decisive correlations remain local.
A recent DCA study of the two-dimensional extended Hubbard model suggested that this assumption is not justified outside the charge-ordered
phase.~\cite{PhysRevB.95.115149} Further development of EDMFT is nevertheless interesting for applications away from the charge-order,
since it can be used as a stepping stone for nonlocal extensions like the dual boson approach~\cite{Rubtsov20121320} or $GW$
methods.~\cite{PhysRevB.66.085120}

The use of ED in the context of EDMFT is more difficult than in DMFT due to the retarded interaction in the Holstein-Anderson impurity model.
This interaction is mediated by a bosonic bath which is coupled to the charge of the impurity.
In principle, this leads to an infinite size of the impurity Hamiltonian, even for a single bosonic energy level.
In this paper we report on an ED solver for the Holstein-Anderson model which we use to solve the EDMFT equations.
We show on the example of the extended Hubbard model on the square lattice with nearest neighbor interactions
that the proposed numerical scheme reproduces a convincing description of the EDMFT phase diagram.
We benchmark our ED scheme with a CTQMC solver and discuss its applicability in different physical regimes of the extended Hubbard model.
We find that the ED is a recommendable alternative to strong-coupling CTQMC solvers in the deeply insulating phase.

This paper is organized as follows:
In Sec.~\ref{sec:edmft} we briefly recall the EDMFT self-consistency loop. We describe an efficient ED solver for the Holstein-Anderson impurity model in Sec.~\ref{sec:solver}
and use it so solve the EDMFT equations for the extended Hubbard model in Sec.~\ref{sec:application}. We conclude with a discussion of our results in Sec.~\ref{sec:conclusions}.

\section{Extended DMFT}
\label{sec:edmft}
We briefly recollect the EDMFT scheme. To this end, we introduce the extended Hubbard model,
\begin{equation} 
    H_{\text{latt}} = 
    \sum_{ij\sigma} t_{ij}  c_{i\sigma}^{\dagger} c_{j\sigma}  +
	U\sum_{i} n_{i\uparrow}n_{i\downarrow} +\frac{1}{2}\sum_{ij}{V_{ij}}n_{i}n_{j}.
\label{hubbard_hamiltonian}
\end{equation}
Here, $t_{ij}$ is the hopping integral between the sites $i$ and $j$,
$c_{i\sigma}^{\dagger}$ and $c_{i\sigma}$ create or annihilate an electron with spin $\sigma$ on the site $i$, respectively.
$U$ is the on-site Hubbard repulsion, and $V_{ij}$ a nonlocal coupling between the charges $n_i=\sum_{\sigma}c_{i\sigma}^{\dagger}c_{i\sigma}$ on different sites $i$ and $j$.

Within the EDMFT an approximation to the extended Hubbard model is obtained by solving an effective Holstein-Anderson impurity
model:~\cite{0953-8984-14-3-312}
\begin{align}  
    H_{\text{imp}} = \sum_{\sigma}\varepsilon_{d}d^{\dagger}_{\sigma}d_{\sigma} + 
 U n_{d\uparrow} n_{d \downarrow} + &
  \sum_{k\sigma}
  ({\mathcal{V}_{k}} d^{\dagger}_{\sigma}f_{k\sigma}+h.c.)\nonumber \\ 
  +\sum_{k,\sigma}\varepsilon_{k}f^{\dagger}_{k\sigma}f_{k\sigma} +  
  \sum_{p}\Omega_{p}b^{\dagger}_{p}b_{p} + 
   \sum_{p} &
  {\mathcal{W}_{p}}(b^{\dagger}_{p}+b_{p}){\bar{n}_d}. 
\label{imp}
\end{align}
In this model $\varepsilon_{d}$ and $\varepsilon_{k}$ are the energy levels of the impurity and of the fermionic bath, respectively.
The operator $d^{\dagger}_{\sigma}$ ($d_{\sigma}$) creates (annihilates) a fermion with spin $\sigma$ on the impurity, $f^{\dagger}_{k\sigma}$ ($f_{k\sigma}$) are fermionic bath operators.
$n_{d\sigma}=d^{\dagger}_{\sigma}d_{\sigma}$ is the particle number operator of the impurity,
$\mathcal{V}_{k}$ is the hybridization between impurity and fermionic bath level with index $k$.
The impurity is further coupled to the bosonic bath with energy levels $\Omega_{p}$.
The operator $b^{\dagger}_{p}(b_{p})$ creates (annihilates) a boson in the energy level with index $p$,
$\mathcal{W}_{p}$ describes the coupling between bosons and the fluctuation of the impurity's charge density $\bar{n}_d = \sum_\sigma(n_{d\sigma} - \langle n_{d\sigma} \rangle)$.

\paragraph*{\textbf{Self-consistent cycle.}}

In EDMFT the energy levels $\varepsilon_{k}$ and $\Omega_{p}$ of the fermionic and bosonic baths in Eq.~\eqref{imp},
as well as the couplings $\mathcal{V}_{k}$ and $\mathcal{W}_{p}$ with the impurity, have to be fixed self-consistently.
To do this, one needs to express the local correlation functions of the lattice in terms of the corresponding correlation functions of the Holstein-Anderson model.
We denote by $G_{\text{loc},\sigma}(i\omega)=-\langle c_\sigma c^\dagger_\sigma\rangle_\omega$ the local lattice Green's function
and by $X_{\text{loc}}(i\nu)=-\langle \bar{n}\bar{n}\rangle_\nu$ the local part of the (connected) charge susceptibility of the extended Hubbard model, Eq.~\eqref{hubbard_hamiltonian}.
Here, $c^\dagger_\sigma$ ($c_\sigma$) and $n=\sum_\sigma c^\dagger_\sigma c_\sigma$ ($\bar{n}=n-\langle n\rangle$) are the creation (annihilation)  operator and the charge density (fluctuation) of the lattice on a given site.
$\omega_m = 2(m+1)\pi/\beta$ and $\nu_m = 2m\pi/\beta$ are the fermionic and bosonic Matsubara frequencies, respectively. $\beta$ is the inverse temperature.

On the other hand, we define the correlation functions of the Holstein-Anderson Hamiltonian, Eq.~\eqref{imp}, as $G_\sigma(i\omega)=-\langle
d_\sigma d^\dagger_\sigma\rangle_{\textit{imp},\omega}$ and $X(i\nu)=-\langle \bar{n}_d\bar{n}_d\rangle_{\textit{imp},\nu}$, where the label
``imp'' denotes a thermal average in the impurity model. The spin index $\sigma$ on Green's function will be suppressed in the following,
since we consider only paramagnetic solutions of the extended Hubbard model.
With the above definitions, the EDMFT equations read,
\begin{subequations}
\begin{align}
    G_{\text{loc}}(i\omega_n)&=\frac{1}{N}\sum_{\kv}\frac{1}{G^{-1}(i\omega_n)+\Delta(i\omega_n)-t(\kv)}\label{eq:edmft_g},\\
    X_{\text{loc}}(i\nu_n)&=\frac{1}{N}\sum_{\qv}\frac{1}{X^{-1}(i\nu_n)+\Lambda(i\nu_n)-V(\qv)}\label{eq:edmft_x}.
\end{align}
\label{eq:edmft_sc} 
\end{subequations}
Here we have introduced the Fourier transforms $t(\kv)$ and $V(\qv)$ of the electronic hopping and of the nonlocal interaction potential of
the extended Hubbard model, Eq.~\eqref{hubbard_hamiltonian}, respectively. $N$ is the number of lattice sites. We refer to	$\Delta
(i\omega_n) = \sum_{k}{|\mathcal{V}_{k}|}^2/(i\omega_n - \varepsilon_k)$ 
and $\Lambda (i\nu_n) = \sum_{p}2|\mathcal{W}_{p}|^2 \Omega_{p}/[(i\nu_n)^2 -
\Omega^2_p]$ as the fermionic and bosonic hybridization functions, respectively.

To solve the EDMFT equations~\eqref{eq:edmft_sc} self-consistently, one needs to adjust the hybridization functions $\Delta, \Lambda$.
To this end, one starts with an initial guess for $\Delta, \Lambda$ and solves the impurity model, Eq.~\eqref{imp}.
New hybridization functions are obtained by update formulae, such as
\begin{subequations} 
\begin{align}
    \Delta_{\text{new}}(i\omega_n)=&\Delta_{\text{old}}(i\omega_n)\hspace{-.05cm}+\hspace{-.05cm}\xi[G^{-1}(i\omega_n)\hspace{-.05cm}-\hspace{-.05cm}G^{-1}_{\text{loc}}(i\omega_n)],\\
    \Lambda_{\text{new}}(i\nu_n)  =&\Lambda_{\text{old}}(i\nu_n)\hspace{-.05cm}+\hspace{-.05cm}\xi[X^{-1}(i\nu_n)\hspace{-.05cm}-\hspace{-.05cm}X^{-1}_{\text{loc}}(i\nu_n)],\label{eq:lambda_update}
\end{align}
\label{eq:delta_update}
\end{subequations}
where $0<\xi\leq1$ is a dimensionless mixing parameter to control convergence. The impurity model is solved in turn, this process is repeated until
convergence, that is, when $G(i\omega_n)=G_{\text{loc}}(i\omega_n)$ and $X(i\nu_n)=X_{\text{loc}}(i\nu_n)$.

\section{Impurity solver}
\label{sec:solver}
In the following we will describe the numerical implementation of a solver for the Holstein-Anderson model, Eq.~\eqref{imp}, based on the exact diagonalization approach.
The basic advantages of the ED~\cite{PhysRevLett.72.1545, S0217984994000571} compared to quantum Monte Carlo methods
are the applicability at very low temperatures and the possibility to calculate correlation functions on the real frequency axis, free of statistical noise.
In order to calculate the impurity correlation functions by means of ED, one needs to find the eigenvalues and eigenvectors of the impurity Hamiltonian Eq.~\eqref{imp}.
The Hilbert space of the bosons mediating the retarded interaction $\Lambda$ is infinite and thus needs to be truncated in order to obtain a finite Hamiltonian.
This corresponds to a maximal number of bosons that can be excited at the same time,
one has to make sure that the average number of bosons $\langle b^{\dagger}_{p}b_{p}\rangle$ is smaller than this cutoff for each bosonic bath level $p$.
Moreover, the fermionic and bosonic hybridization functions $\Delta$ and $\Lambda$ have to be discretized.
A projection method from a continuous to a discrete hybridization function $\Delta$ was
described by Caffarel et al~\cite{PhysRevLett.72.1545}. In the same spirit, we project the continuous fermionic and bosonic hybridization functions
$\Delta$ and $\Lambda$ to discretized ones by limiting ourselves to a number of $K$ fermionic and $P$ bosonic bath levels.
The discretized hybridization functions are then given as
\begin{subequations}
\begin{align}
    \Delta (\omega_n) &\approx \Delta^{K} (\omega_n) = \sum_{k=1}^K\frac{{|\mathcal{V}_{k}|}^2}{i\omega_n - \varepsilon_k},\\
    \Lambda (\nu_n) &\approx \Lambda^{P} (\nu_n) = \sum_{p=1}^P\frac{2|\mathcal{W}_{p}|^2 \Omega_{p}}{(i\nu_n)^2 -
    \Omega^2_p}\label{eq:lambda}.
\end{align}
\label{eq:min}
\end{subequations}
When solving the EDMFT equations~\eqref{eq:edmft_sc} the bath discretization is enforced as follows:
One first solves the impurity problem with an intial guess for the bath parameters $\mathcal{V}_{k}$, $\varepsilon_k$, $\mathcal{W}_p$, and $\Omega_{p}$ in Eq.~\eqref{eq:min},
corresponding to discrete hybridization functions $\Delta^K_\text{old}$ and $\Lambda^P_\text{old}$.
After that one obtains new hybridization functions $\Delta_{\text{new}}$ and $\Lambda_{\text{new}}$ from the update formula Eq.~\eqref{eq:delta_update}.
In general, it is not possible to express $\Delta_{\text{new}}$ and $\Lambda_{\text{new}}$ exactly using only $K$ fermionic and $P$ bosonic bath levels.
In order to obtain approximate new hybridization functions $\Delta^K_{\text{new}}$ and $\Lambda^P_{\text{new}}$,
that can be expressed via a set of new bath parameters in Eq.~\eqref{eq:min},
we use a non-linear least-squares method that minimizes the following expressions,
\begin{subequations}
\begin{align}
    \chi^2_F = & \frac{1}{N_{\omega}+1}\sum_n^{N_{\omega}}\left[\Delta_{\text{new}}(\omega_n) - \Delta^K_\text{new}(\omega_n)\right]^2, \\
    \chi^2_B = & \frac{1}{N_{\nu}+1}\sum_n^{N_{\nu}}\left[\Lambda_{\text{new}}(\nu_n) - \Lambda^P_\text{new}(\nu_n)\right]^2.
\end{align}
    \label{eq:leastsqr}
\end{subequations}
Here, $N_{\omega}$ and $N_\nu$ are the numbers of fermionic and bosonic Matsubara frequencies for which the least-squares fit is performed.
The minimization of the expressions in Eq.~\eqref{eq:leastsqr} is achieved by adjusting the bath parameters on the RHS of Eq.~\eqref{eq:min}.
This yields new \textit{discrete} hybridization functions $\Delta^K_{\text{new}}$ and $\Lambda^P_{\text{new}}$,
which are then used in the next iteration of the EDMFT cycle. This process is repeated until convergence.

We note that due to the bath discretization the performance of the update formula depends on the physical regime.
The formula given in Eq.~\eqref{eq:lambda_update} accelerates the fit of $\Lambda$ at high energies and is useful in insulating
regimes. In metallic regimes a fast convergence at low energies is more important and one may use the following update formula instead,
\begin{subequations}
\begin{align}
    \Lambda_{\text{new}}(i\nu_n)=&\Lambda_{\text{old}}(i\nu_n)+\zeta[X_{\text{loc}}(i\nu_n) - X(i\nu_n)],
\end{align}
\label{eq:delta_update_2}
\end{subequations}
where $\zeta$ is a mixing parameter of suitable dimension.

\textbf{\textit{Bath representation}}.
We continue with a description of the construction of the Hilbert space of the Holstein-Anderson impurity, Eq.~\eqref{imp}, which is coupled to a fermionic and a bosonic bath.
The procedure of constructing the fermionic Hilbert space was already described in previous works, e.g., Ref.~\onlinecite{Lin_Gubernatis}.
We will therefore concentrate on the details of representing the bosonic bath.
In order to truncate the infinite Hilbert space of the bosons, an occupation number cutoff $N_p$ for each bosonic bath level $p$ is introduced.
Then, one has for the occupation number $n_{bp}$ of the bosons in the bosonic bath level $p$: $0\leq n_{bp}\leq N_p$.
Likewise, we label as $n_{fk}$ the fermionic occupation number in the fermionic bath level $k$, which is either $0$ or $1$.
We denote the occupation number of the impurity level as $n_d$. With these definitions we express the basis states $\ket{\psi}$ of the Holstein-Anderson model as follows,
\begin{align}
    \nonumber\ket{\psi} =&\underbrace{| n_{b1}, n_{b2}, \ldots, n_{bp},...\rangle}_{\text{boson part}} \\
                  \otimes& \underbrace{| n^{\uparrow}_{d}, n^{\uparrow}_{f1}, \ldots, n^{\uparrow}_{fk}\ldots, n^{\downarrow}_{d}, n^{\downarrow}_{f1}, \ldots, n^{\downarrow}_{fk},\ldots\rangle}_{\text{fermion part}}.
                \label{base_vector}
\end{align}
For a total number of $P$ bosonic bath levels the basis of bosonic states has a dimension $(N_1+1)(N_2+1)\cdots(N_P+1)$.
For $K$ fermionic bath levels and taking a spin-multiplicity of $2$ into account the fermionic basis (of one impurity and $K$ bath states) is of dimension $(2\cdot 2^K)^2$.
The size of the truncated Holstein-Anderson Hamiltonian, Eq.~\eqref{imp}, is hence $[2^{2K+2}(N_1+1)(N_2+1)\cdots(N_P+1)]^2$.
However, due to conservation of the total particle number and spin, the Hamiltonian decomposes into blocks,
with the largest one being of size $[(K+1)!(\frac{K+1}{2})!^{-2}]^2[(N_1+1)(N_2+1)\cdots(N_P+1)]^2$.~\cite{Lin_Gubernatis}
A feasible example, that was used in our benchmarks with a CTQMC solver, is $K=7$, $P=3$, and $N_p=7$,
leading to a size of $\sim 2.5\times10^6$ elements in the largest block and a total memory requirement of $\sim1$ - $1.5$Gb.

\textbf{\textit{Implementation details.}} 
The Hamiltonian matrix generated from the Holstein-Anderson Hamiltonian and the basis states, Eq.~\eqref{base_vector}, is a sparse Hermitian
matrix. For calculations at not too high temperatures we need only the eigenstates corresponding to the lowest energies.
Avoiding matrix-matrix multiplications, the lowest eigenvalues and corresponding eigenstates can be found by the Arnoldi
method~\cite{arpack}, based on the Krylov subspaces.~\cite{Krylov}

\textbf{\textit{Correlation functions.}}
Having access to the eigenvalues $E_{l}$ and the corresponding eigenvectors $|l\rangle$ of the Holstein-Anderson Hamiltonian,
one calculates Green's function at finite temperature from the Lehmann expression,
\begin{align}
	G_\sigma (z) = 
    \frac{1}{\mathcal{Z}} \sum_{ll'} {\frac{|\bra{l'} d^{\dagger}_\sigma\ket{l}|^2}{z + E_{l} - E_{l'}}}\left( e^{-\beta E_l} + e^{-\beta E_{l'}}  \right),
    \label{one-particleGF}
\end{align}
where $\mathcal{Z} = \sum_{l}e^{-\beta E_l}$ is the partition function and $z\in\mathbb{C}$ a point in the complex plane. Similarly, one calculates the charge susceptibility,
\begin{align}
	X (z') =
    -\frac{1}{\mathcal{Z}} \sum_{ll'}{\frac{\left| \bra{l'}\bar{n}_d \ket{l}\right|^2}{z' + E_l - E_{l'}}} \left( e^{-\beta E_l} - e^{-\beta E_{l'}} \right).
   \label{two-particleGF}
\end{align}
At low temperatures the exponentials in Eqs.~\eqref{one-particleGF} and~\eqref{two-particleGF} become very small for high energies.
Typically, the larger the expectation value $\bra{l}b^{\dagger}_pb_p\ket{l}$ of the bosonic occupation number is in an eigenstate $\ket{l}$, the higher is the energy $E_l$ of this state.
The error introduced due to the truncation of the bosonic Hilbert space therefore becomes exponentially smaller with the cutoff value $N_p$ of the bosonic occupation number.

In ED one straightforwardly calculates the correlation functions on the real axis, i.e., $z\rightarrow\omega+i0^+$ and $z'\rightarrow\nu+i0^+$ in Eqs.~\eqref{one-particleGF} and~\eqref{two-particleGF}, respectively.
However, for a comparison of results with CTQMC codes it is convenient to calculate the correlation functions at the Matsubara energies $i\omega_m$ and $i\nu_m$, respectively.
In a numerical implementation this leads to a division by zero at the bosonic Matsubara energy $z'\rightarrow i\nu_0=0$ in Eq.~\eqref{two-particleGF} for degenerate energies $E_l$ and $E_{l'}$.
In the appendix we propose a way to calculate a bosonic function at $i\nu_0=0$ without having to evaluate Eq.~\eqref{two-particleGF} at this point.
This requires less implementation effort than the separate treatment of degenerate and non-degenerate energies.
\begin{figure}[htp!]
\includegraphics[width=0.48\textwidth]{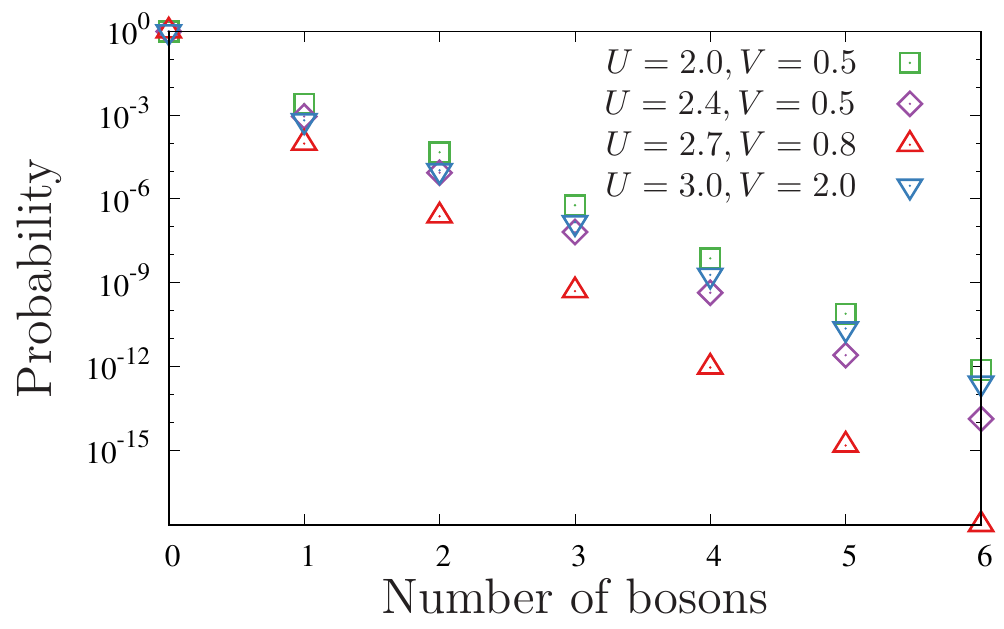}
    \caption{(Color online)
    Discrete probability distribution of the total number of bosons.
    Each data point corresponds to the probability of finding the respective number of bosons excited.
    The data points agree very well with an exponential decay for all considered parameters.
}
\label{fig:bosonic_distribution}
\end{figure}

\begin{figure*}[t]
        \begin{center}
        \vspace{-0.5cm}
        \large
        $U = 1$, $V = 0.15$
        \end{center}

        \includegraphics[width=0.48\linewidth]{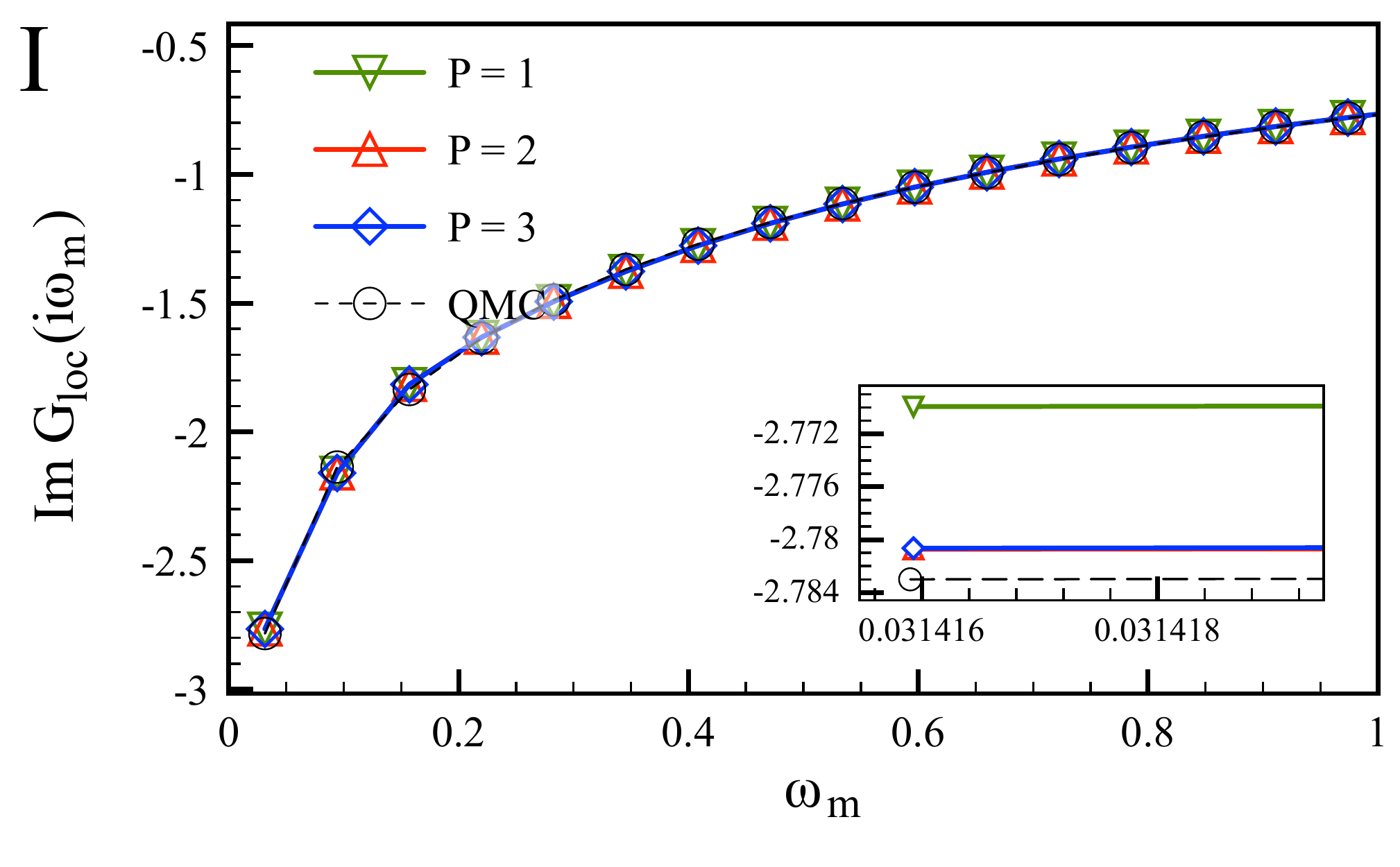}
        \hspace{.5cm}
        \includegraphics[width=0.48\linewidth]{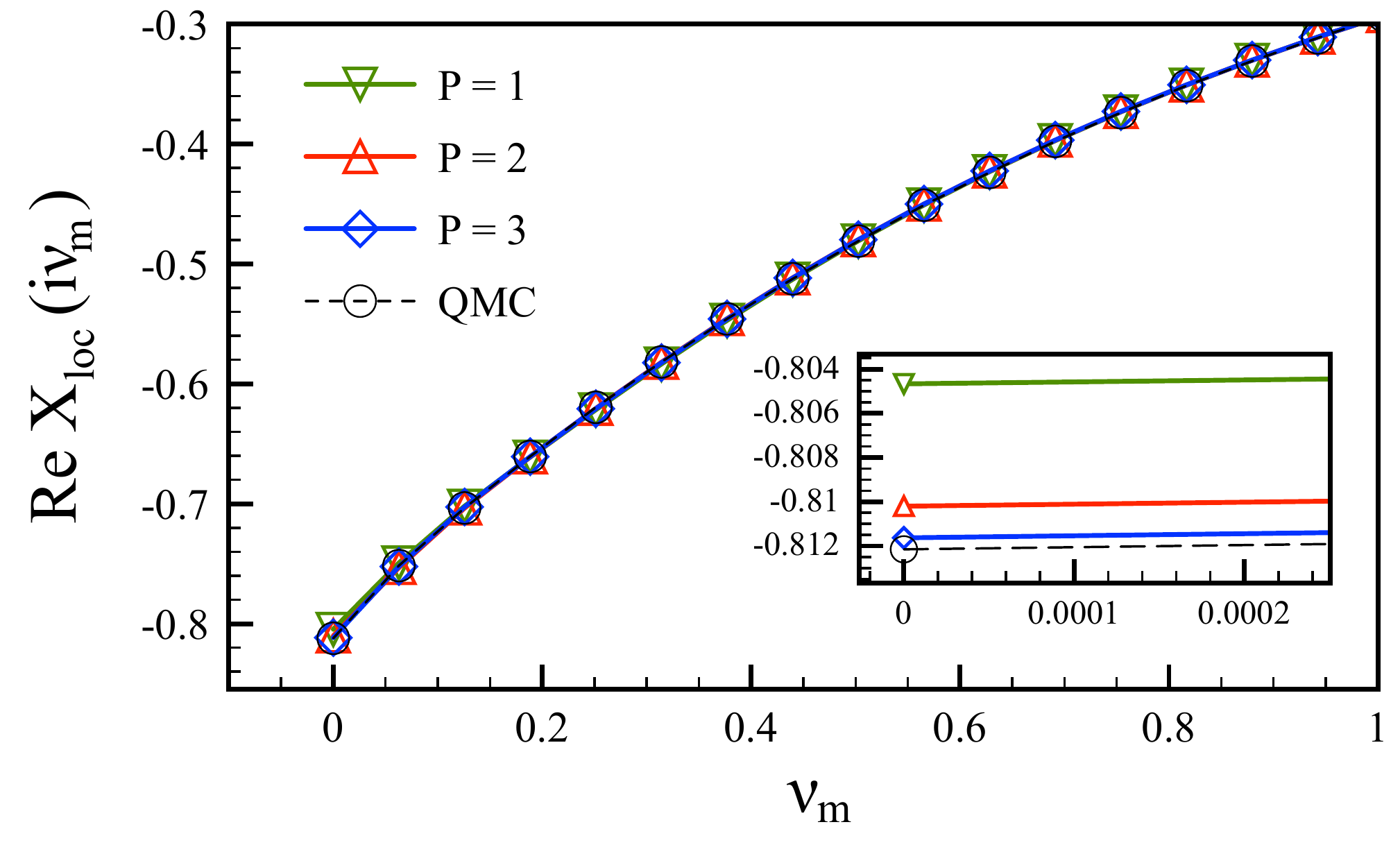}
        \vspace{-0.2cm}
        \begin{center}
        \vspace{-0.5cm}
        \large
        $U = 2$, $V = 0.5$
        \end{center}

        \includegraphics[width=0.48\linewidth]{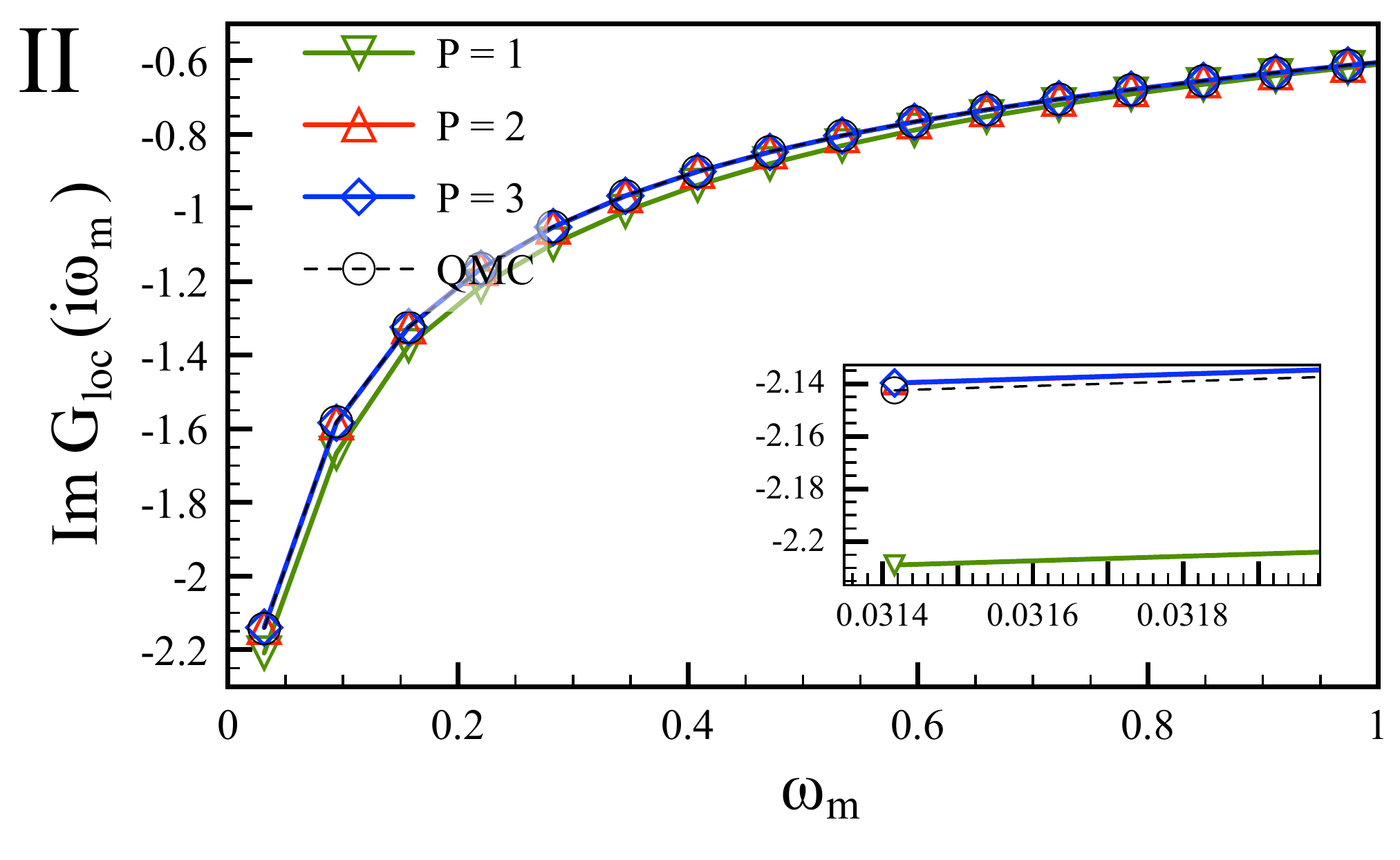}
        \hspace{.5cm}
        \includegraphics[width=0.48\linewidth]{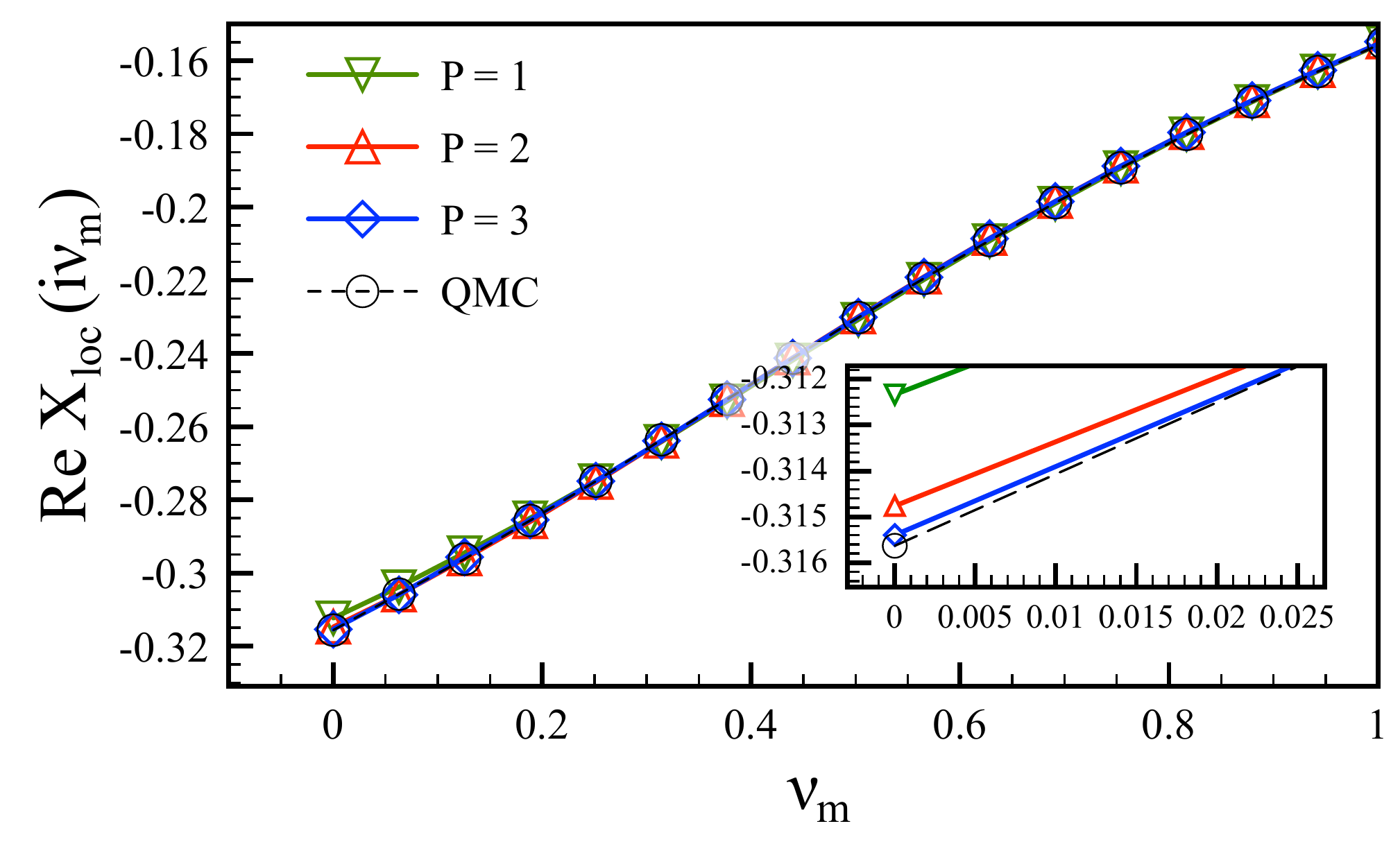}

        \vspace{-0.2cm}
        \caption{(Color online) Benchmarks of ED with strong-coupling CTQMC in the weakly (top) and moderately (bottom) correlated FL (see also points I and II in Fig.~\ref{PDV}, respectively).
            The figures show the self-consistent local correlation functions of EDMFT on the Matsubara axis [cf. definitions~\eqref{eq:edmft_g} and~\eqref{eq:edmft_x}].
            Results from ED (triangles and squares) with $P=1,2,3$ bosonic bath levels and from CTQMC (circles) are shown. 
            Insets show the lowest Matsubara frequencies $\omega_0$ and $\nu_0$. There, the size of the circles indicates the statistical error of the CTQMC results.
    }
        \label{weak}
\end{figure*}

\section{Results}
\label{sec:application}
    We discuss the applicability of our ED solver in different physical regimes and perform a benchmark with a strong-coupling CTQMC solver.
We calculate the EDMFT phase diagram of the 2D extended Hubbard model with nearest neighbor interactions, Eq.~\eqref{hubbard_hamiltonian}.
Real axis spectra are shown and we discuss the effects of screening in the metallic and insulating phases.
In all benchmarks and applications of this section the nearest neighbor hopping is set to $t = 0.25$ and the inverse temperature to $\beta = 100$.
\subsection{Benchmark with strong-coupling CTQMC}
\label{sec:benchmarks}
We evaluate the effects of the bath discretizations and of the Hilbert space truncation in our ED solver.
To this end, let us consider the different simplifications that are necessary to make the Holstein-Anderson Hamiltonian, Eq.~\eqref{imp},
tractable within the ED.

\textit{Firstly}, we have limited the number $K$ of fermionic bath energies $\varepsilon_k$.
In this respect our ED scheme behaves similarly to ED solvers for the DMFT equations.
For details on the effects of the fermionic bath discretization we refer the reader to the literature referenced in the introduction.
Typically, more fermionic bath levels are needed close to the metal-insulator transition, the resolution of this transition by means of ED
is challenging.~\cite{PhysRevLett.72.1545} In practice, we used a fixed number of $K=7$ fermionic bath sites for our benchmarks with the
CTQMC solver and $K=5$ for the calculation of the EDMFT phase diagram. Due to the large Hilbert space of the bosonic system, $K\sim10$ is
the limit where matrix sizes remain manageable.

\textit{Secondly}, we have limited the maximal occupation number $N_p$ of the bosons that can be excited in the energy level $\Omega_p$.
In order to validate that this truncation is justified, we computed the probability distribution of the average number of bosons,
$N_{\text{tot}}=\frac{1}{P}\sum_{p=1}^P\langle b^\dagger_pb_p\rangle$. This is shown in Fig.~\ref{fig:bosonic_distribution} for several parameter regimes.
It appears that in the parameter regimes considered in our calculations the average number of bosons is very small and that the probability
that a certain number of bosons is excited decays exponentially. This may be explained by the fact that each additional boson increases the
energy by approximately $\Omega_p$, which enters thermal averages via an additional factor $\sim e^{-\beta\Omega_p}$.

While the average number of bosons in low-energy eigenstates is always small at the considered temperatures,
larger bosonic occupancies contribute to eigenstates with higher energies.
When calculating static averages of the form $\sum_l\langle l|\cdots|l\rangle e^{-\beta E_l}$, where $|l\rangle$ is an eigenstate of the impurity Hamiltonian,
high energy eigenstates may be neglected at the low temperature considered here. In case of dynamical correlation functions the situation is different.
The matrix elements $\langle l|\cdots|l'\rangle$ in the Lehmann expressions in Eqs.~\eqref{one-particleGF} and~\eqref{two-particleGF}
introduce non-negligible overlaps of low and high energy eigenstates, where the latter may correspond to larger bosonic occupancies.
Hence, while a small cutoff $N_p$ is sufficient for static averages, dynamical correlation functions require a larger $N_p$, in particular at high frequencies.
We observe that for all parameter regimes of our applications it is sufficient to have a maximal occupancy of $N_p=7$ bosons in each energy level.
We note that at high temperatures one can expect the ED to become unfeasible due to a large
bosonic occupancy $\langle b^\dagger_p b_p\rangle\sim [e^{\beta\Omega_p}-1]^{-1}$, which in turn requires an increase in the maximal
occupation number $N_p$.
\begin{figure*}[t]
        \begin{center}
        \large
        $U = 2.4$, $V = 0.5$
        \end{center}

        \includegraphics[width=0.48\linewidth]{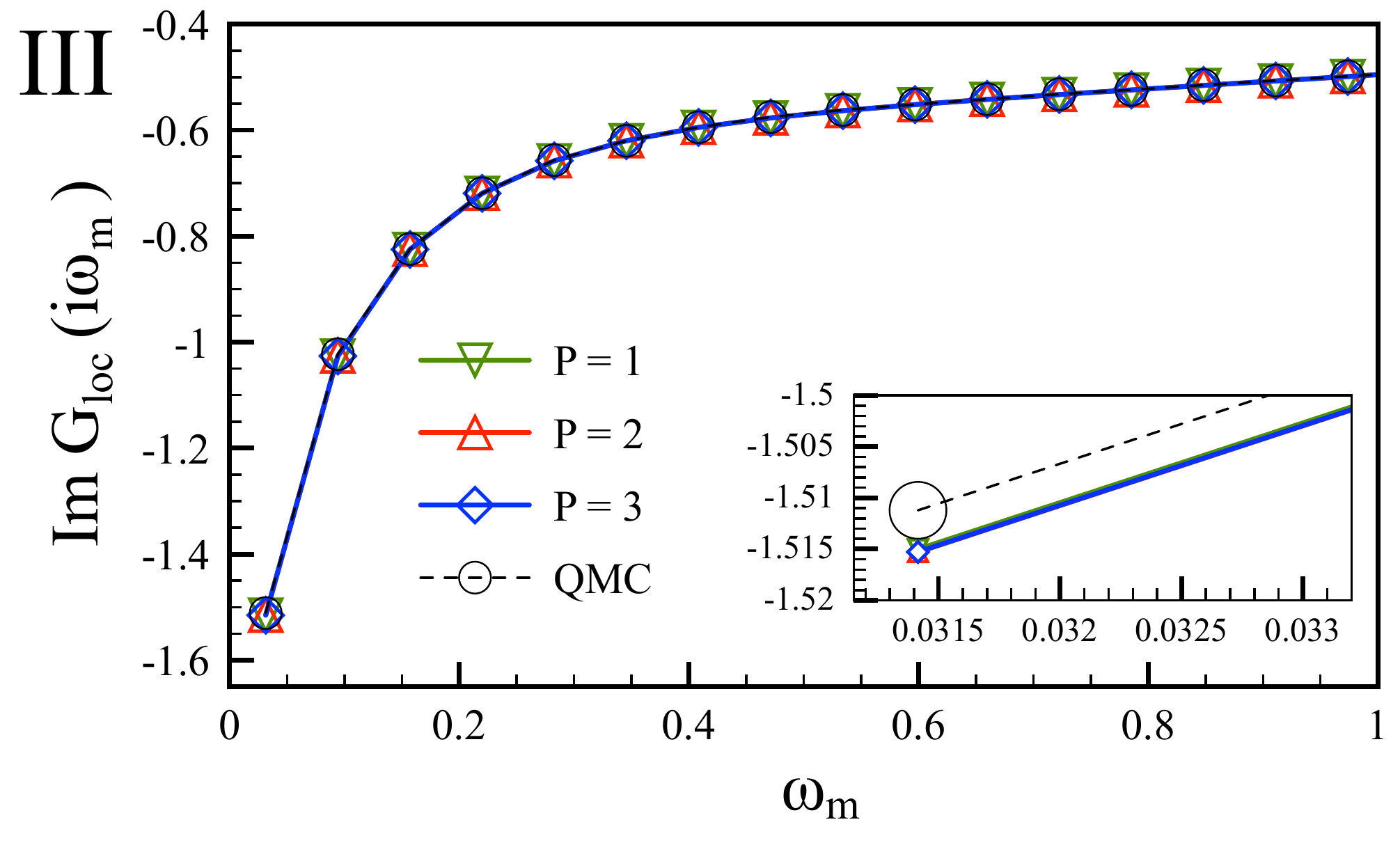}
        \hspace{.5cm}
        \includegraphics[width=0.48\linewidth]{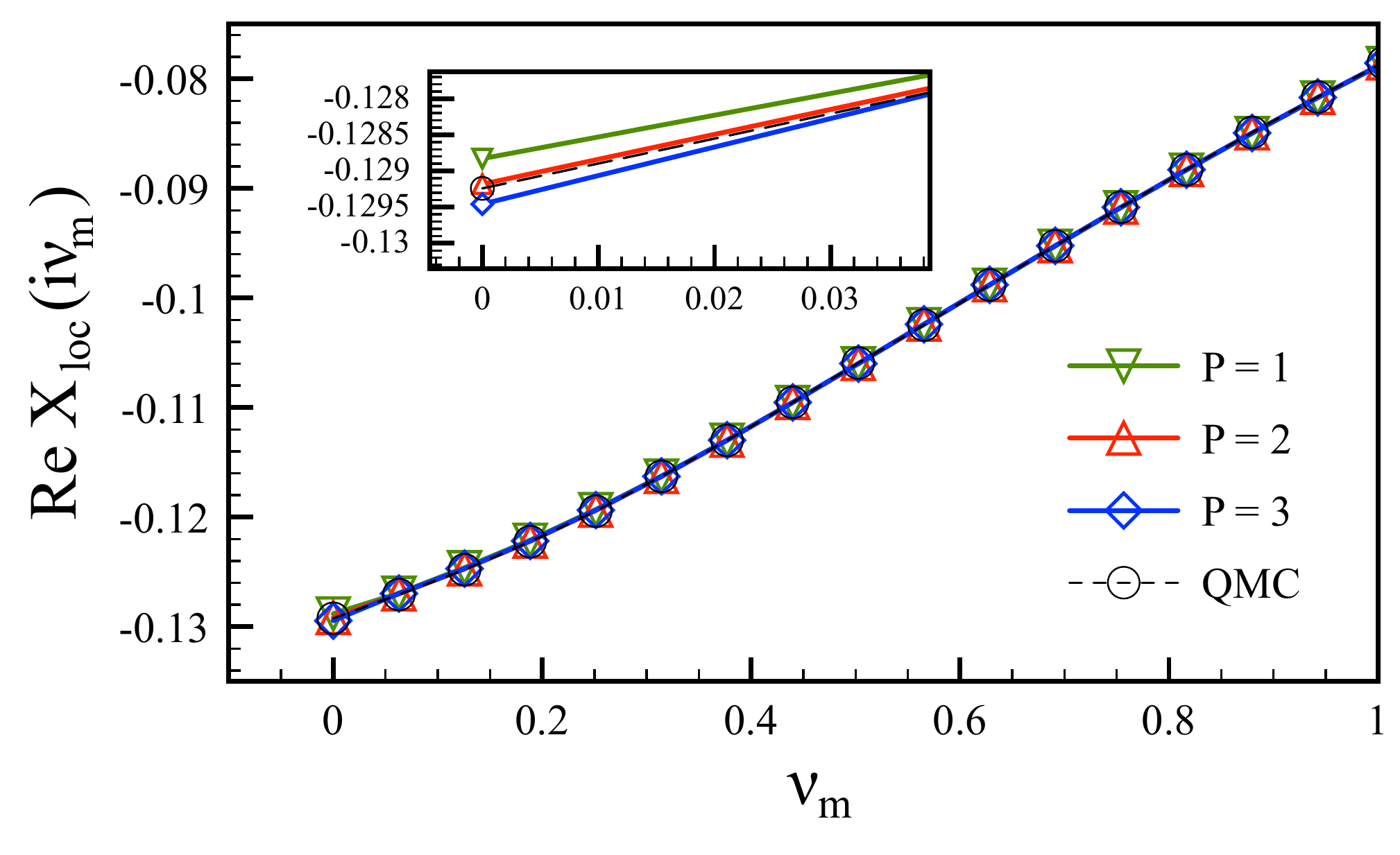}
        \begin{center}
        \large
        $U = 2.7$, $V = 0.8$
        \end{center}

        \includegraphics[width=0.48\linewidth]{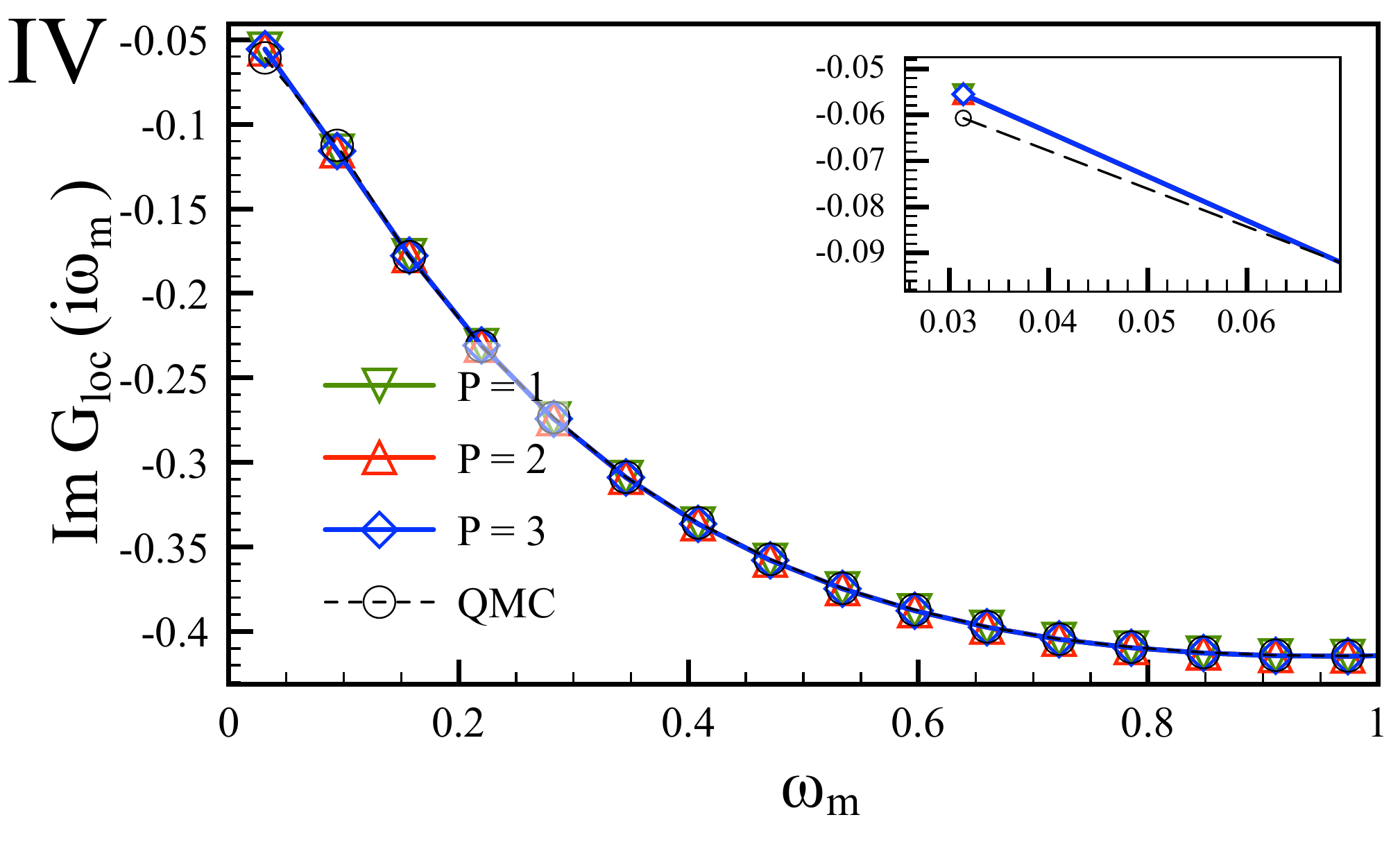}
        \hspace{.5cm}
        \includegraphics[width=0.48\linewidth]{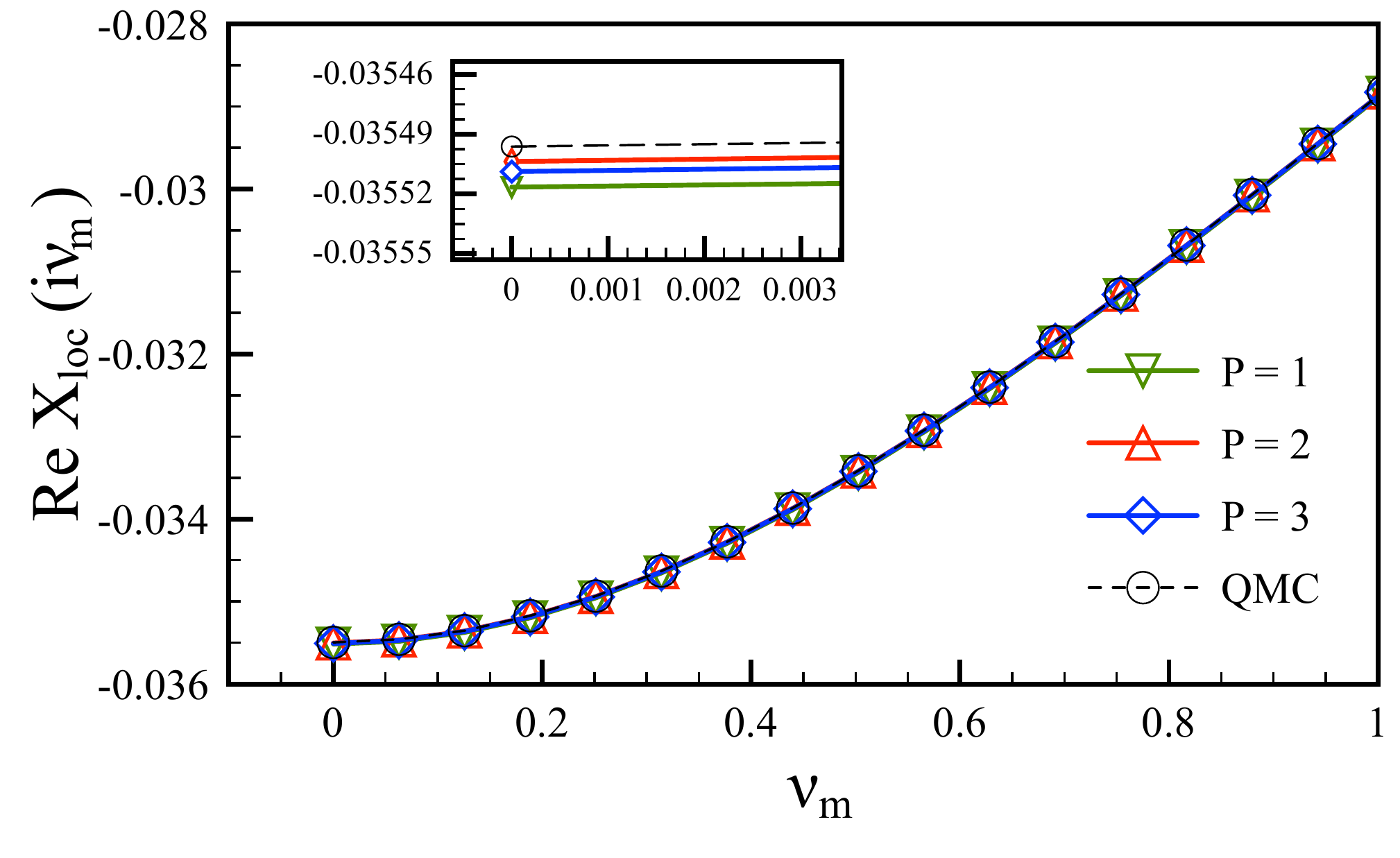}
        \caption{
            (Color online)
            Benchmarks of ED with strong-coupling CTQMC in the strongly correlated FL (top) and in the MI (bottom).
            See also the respective points III and IV in Fig.~\ref{PDV}. Conventions as in Fig.~\ref{weak}.
}
    \label{strong}
\end{figure*}

\begin{figure}
	\includegraphics[width=0.98\linewidth]{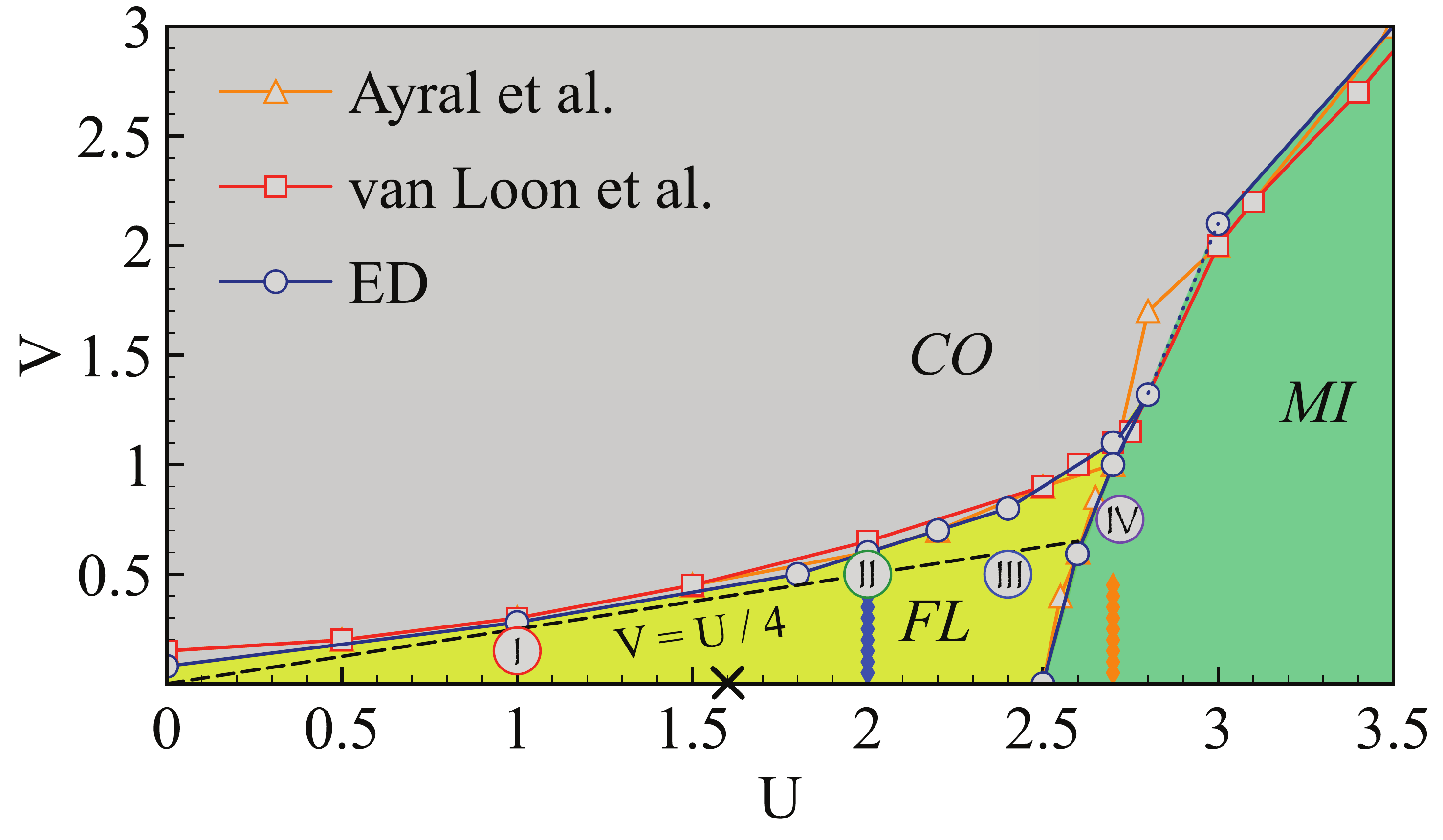}
    \caption{\label{PDV}(Color online) EDMFT phase diagram of the extended Hubbard model in the $U$-$V$ plane.
        Circles indicate the phase boundary obtained by ED.
        Triangles and squares mark the phase boundaries from references \onlinecite{PhysRevB.87.125149} and \onlinecite{PhysRevB.90.235135}, respectively.
        Lines between data points are guides to the eye.
        Roman numerals indicate where benchmarks with CTQMC were performed (see Sec.~\ref{sec:benchmarks}).
        The dashed black line at $V=U/4$ is a rough estimate of the FL-CO transition.
        Blue and yellow dots mark two series of calculations in the FL and MI discussed in Sec.~\ref{sec:real_axis}.
        The black cross lies at the FL-MI transition of the 2D Hubbard model predicted by the DCA~\cite{PhysRevLett.110.216405}.}
\end{figure}

\textit{Thirdly}, we have limited the number $P$ of bosonic energy levels $\Omega_p$.
It is a priori not clear how $P$ should be chosen and if it is feasible for ED calculations to reach a convergence in $P$.
Here we rely on a benchmark with strong-coupling CTQMC, which is not limited in the number of bosonic bath levels.
We solved the EDMFT equations with our ED solver for $P=1,2,3$ in several physical regimes of the extended Hubbard model (compare Roman
numerals I-IV in the phase diagram Fig.~\ref{PDV} in the following subsection).
We then compared Green's function and the susceptibility on the Matsubara axis to solutions obtained from the CTQMC solver presented
in Ref.~\onlinecite{Hafermann13}.

The benchmarks were performed in the weakly correlated Fermi Liquid (FL, I) with small $U$ and $V$,
the moderately and strongly correlated FL (II and III) at intermediate $U$ and small $V$, and the Mott insulating (MI) regime (IV), close to the transition to the FL.
According to Fig.~\ref{weak}, in the weakly and moderately correlated FL there is a convergence trend to the CTQMC results with an increasing number $P$ of bosonic bath levels.
The insets in the left and right panels of Fig.~\ref{weak} show that at $P=3$ Green's function and the susceptibility lie very close or within the uncertainty of the CTQMC results.
Likewise, the top panels of Fig.~\ref{strong} show that in the strongly correlated FL (III) $P=2$ to $P=3$ energy levels are sufficient for convergence. 
In the MI phase $P=2$ levels are sufficient (see bottom panels).

\begin{table}
    \caption{Energies $\Omega_p$ and coupling constants $\mathcal{W}_{p}$ of self-consistent EDMFT calculations with different numbers $P=1,2,3$ of bosonic bath levels.
    Roman numerals indicate locations in the phase diagram, see Fig.~\ref{PDV}. Bath parameters $\Omega_p$ and $\mathcal{W}_{p}$ are defined as in Eq.~\eqref{imp}.}
    \begin{tabular}{|c|c|c|cc|ccc|c|}
        \hline
         & & $P=1$ & \multicolumn{2}{c|}{$P=2$} & \multicolumn{3}{c|}{$P=3$} & \\
         & $p$ & $1$ & $1$ & $2$ & $1$ & $2$ & $3$ & \\
        \hline
        U = 1    & $\Omega_p$   & 0.857   & 1.103 & 0.222         & 1.275 & 0.459 & 0.067                &\multirow{2}{*}{I}\\
        V = 0.15 & $\mathcal{W}_{p}$    & 0.165   & 0.161 & 0.0525                & 0.149 & 0.082 & 0.019                   & \\
        \hline
        U = 2 &  $\Omega_p$     & 1.126   &  1.462 & 0.447        &         1.704 & 0.717 & 0.107         &\multirow{2}{*}{II}\\
    V = 0.5   & $\mathcal{W}_{p}$           & 0.407   &  0.385 & 0.158         & 0.342 & 0.238 & 0.038                    & \\
        \hline
        U = 2.4& $\Omega_p$& 1.464        &  1.900 & 0.697        & 2.117 & 0.930 & 0.127               &\multirow{2}{*}{III}\\
       V = 0.5 & $\mathcal{W}_{p}$      & 0.298        &  0.277 & 0.128         &  0.253 & 0.171 & 0.019                     & \\
        \hline
        U = 2.7  &$\Omega_p$    & 2.263        & 2.709 & 1.421         & 3.098 & 1.946 & 0.979          &\multirow{2}{*}{IV}\\
        V = 0.8  & $\mathcal{W}_{p}$            & 0.319        & 0.290 & 0.144   & 0.229 & 0.222 & 0.057                    & \\
        \hline
    \end{tabular}
    \label{table:P123}
\end{table}

In general, going from $P=1$ to $2$ leads to an improvement, whereas $P=3$ bosonic bath levels are not needed in most cases.
This can also be seen from Table~\ref{table:P123}, where we have listed the bath parameters of ED calculations at the benchmark points I-IV of Figs.~\ref{weak} and~\ref{strong}.
At $P=3$ the coupling constant $\mathcal{W}_{3}$ corresponding to the smallest energy $\Omega_3$ becomes very small.
The contribution $\sim\mathcal{W}^2_{3}/\Omega_3$ of this bath level to the bosonic hybridization function $\Lambda$ therefore becomes small as well and $P=2$ should suffice.
On the other hand, as will be discussed later, having at least one low energy mode $\Omega_p$ is important in regimes with a broad spectrum of charge excitations, then $P=1$ is insufficient.
We note that a convergence in the number of bosonic energy levels need not lead to an agreement with the CTQMC data,
since the final accuracy of the ED calculations is limited by the fixed number of $K=7$ fermionic bath levels.

No benchmarks could be performed in the charge-ordered phase (CO), which is not accessible to us because we did not consider sublattices with different fillings,
and in the deep insulator (MI), since we could not obtain reliable results with the strong-coupling CTQMC solver in this regime.
This is due to a large variance in the impurity self-energy $\Sigma$, whose behavior at half-filling and large $U$ resembles that of the atomic limit, $\Sigma(i\omega)\sim 1/i\omega$,
leading to strong noise propagation into the update of the fermionic hybridization function $\Delta$.
We note that our ED scheme converges well in the deeply insulating regime.

In order to determine phase boundaries in ED, convergence in the number of the bosonic bath levels is not always needed to achieve a good agreement with earlier CTQMC results.
We find that for the determination of phase boundaries the number of bosonic bath levels can be chosen as follows:
$P=1$ on the phase boundary between FL and CO. Likewise, $P=1$ between FL and MI, far enough from CO.
$P=2$ in the area where the FL, MI and CO are close (see tip of the FL phase in Fig.~\ref{PDV_close}).
$P=2$ to $P=3$ bosonic energy levels are needed to determine the phase boundary between the MI and CO.
At $P=1$ ED predicts the MI-CO transition at much larger values of $V$ than CTQMC.

%
%
\subsection{U-V phase diagram}
\label{sec:uvdiagram}
Using our ED solver, we calculated the EDMFT phase diagram in the $U$-$V$ plane of the half-filled extended Hubbard model,
Eq.~\eqref{hubbard_hamiltonian}, on the square lattice with nearest neighbor interaction $V$, this can be seen in Fig.~\ref{PDV}.

The observed phases are the Fermi liquid (FL) for not too large $U \lesssim 10t$ and $V \lesssim U / 4$, the Mott insulating phase (MI) for
large $U \gtrsim 10t$ and not too large $V$ and the charge-ordered phase (CO) at large $V$.
The phase boundaries to CO were found by performing calculations close to a divergence of the EDMFT charge susceptibility at $\qv=(\pi,\pi)$ and then increasing $V$.
The transition from FL to MI (and from MI to FL) was roughly estimated from a change in the slope of $G(i\omega)$ at small Matsubara energies
(such as the difference between the top left and bottom left panels of Fig.~\ref{strong}).

For comparison we plotted into Fig.~\ref{PDV} the phase boundaries found by Ayral et al.~\cite{PhysRevB.87.125149} and van Loon et
al.~\cite{PhysRevB.90.235135}, who used strong-coupling CTQMC to solve the effective impurity model.
We observe that in our ED scheme the convergence of the effective fermionic and bosonic baths to a self-consistent solution of the EDMFT
equations becomes slower in the same areas as when using CTQMC.
In general this is close to the phase boundaries.~\cite{PhysRevB.90.235135}

\begin{figure}
	\includegraphics[width=0.98\linewidth]{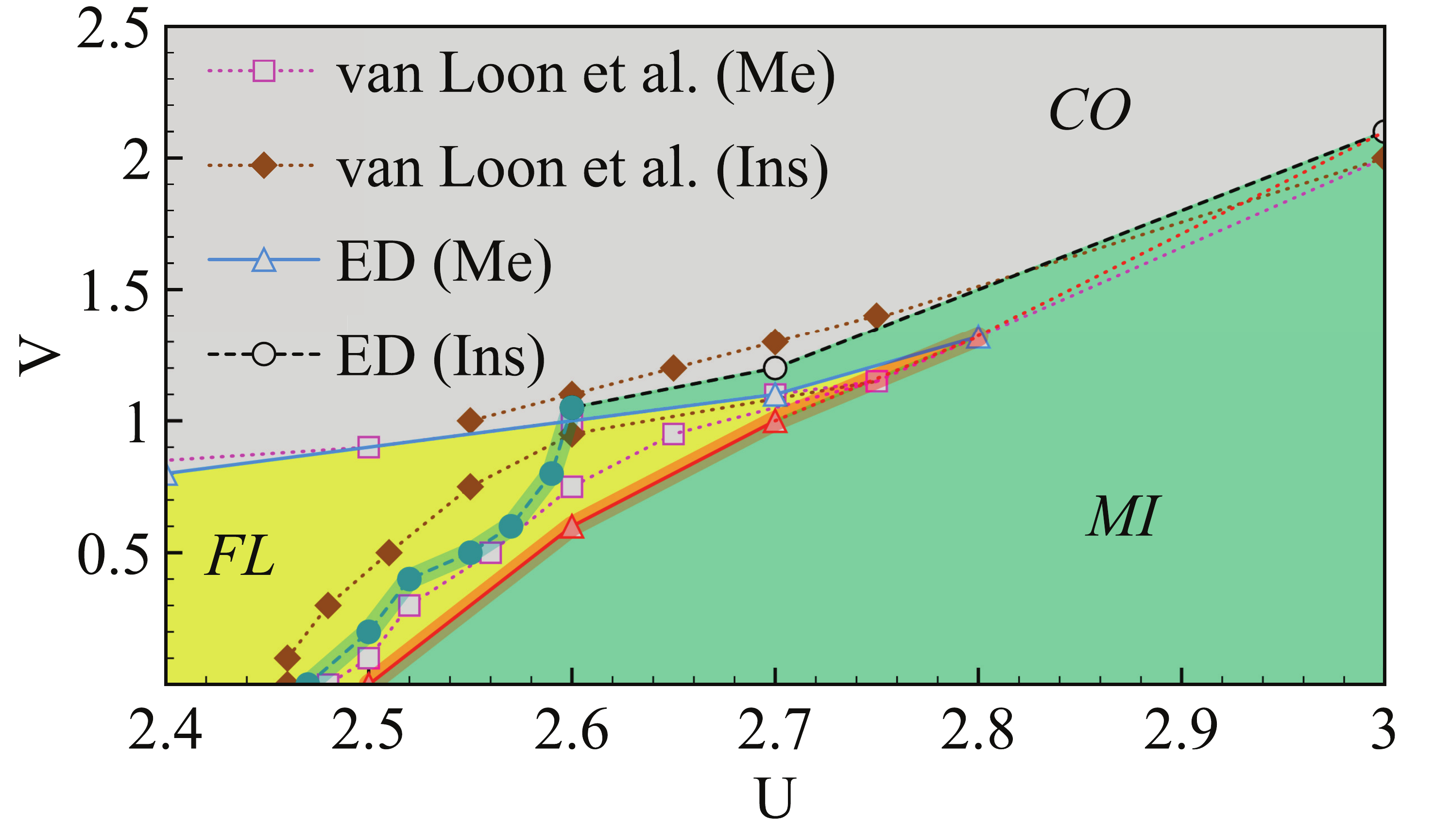}
    \caption{\label{PDV_close}(Color online) Closeup of the FL-MI transition of the EDMFT phase diagram in Fig~\ref{PDV}.
        We compare the phase boundaries predicted by ED (circles and triangles) to data of van Loon et al. (squares and diamonds)~\cite{PhysRevB.90.235135}.
    The FL-MI phase boundary was approached using metallic (triangles) and insulating seeds (circles).
    Compared to van Loon et al., the coexistence region of FL and MI in ED with $K=5$ fermionic bath levels is shifted to slightly larger $U$.
    The shade on the FL-MI phase boundaries indicates the error due to the discretization of the fermionic bath.
}
\end{figure}

We examine more closely the FL-MI phase boundary in Fig.~\ref{PDV_close}: The FL phase is enclosed by the blue phase boundary to CO on the
top and by the thick red boundary to the MI on the right. Coming from the MI, we find the dashed black boundary to CO on the top and the thick green boundary to the FL on the left.
These boundaries reveal the typical coexistence region of the first order FL-MI transition but also a narrow stripe above the yellow region
where MI solutions can be converged while one encounters a transition to CO when coming from the FL. In principle, this can be seen as 
indirect evidence of a MI-CO coexistence but the validation of a first order transition between these phases is not possible since calculations cannot be initiated from charge-ordered seeds.
In general, Figs.~\ref{PDV} and~\ref{PDV_close} show a good quantitative agreement of the phase boundaries with our references.
However, we find that in our ED calculations the coexistence region of FL and MI is shifted to slightly larger values of $U$ compared to the CTQMC results of van Loon et al.
We attribute this shift to the limited number ($K=5$) of fermionic bath levels used in our calculation of the phase diagram (cf. Sec.~\ref{sec:benchmarks}), and the fact that this is an odd number:
The density of states of the half-filled extended Hubbard model on the square lattice is symmetric around the Fermi
level.~\cite{PhysRevB.87.125149} Likewise, the fermionic hybridization function $\Delta(\omega+i\delta)$ is symmetric around $\omega=0$ and hence "gapped" when discretized with an even number of bath levels, favoring the Mott insulator.
On the other hand, an odd number of bath levels favors the FL.
To obtain an estimate of this discretization error we calculated the phase boundary between the FL and MI phase also with an even number $K=4$ of fermionic bath levels.
This estimate is indicated by the line thickness of the green and red FL-MI phase boundaries in Fig.~\ref{PDV_close}.
However, the FL-MI transition at $V=0$ obtained from DCA calculations~\cite{PhysRevLett.110.216405} is at about $U_c \cong 1.6$,
indicating that the error introduced by the EDMFT approximation is much larger than the additional discretization error of ED.

According to these results the ED can be used as a reliable solver for the local reference system of EDMFT, at least at the low temperature considered.
However, the EDMFT approximation as such does not allow for an accurate prediction of the phase boundaries of the extended Hubbard model,
due to the absence of the nonlocal Fock diagram,~\cite{PhysRevB.95.245130} and due to its neglect of nonlocal correlations, that are
relevant in this model.~\cite{PhysRevB.95.115149}
The latter point may be improved upon by considering several sublattices, which also allows to enter the charge ordered phase.
A further option to rectify both of these deficiencies lies in diagrammatic extensions, such as the GW+EDMFT~\cite{Biermann03} or the dual
boson approach,~\cite{Rubtsov20121320,PhysRevB.90.235135} which are appealing fields of application for our ED solver in the future. 

\subsection{Density of states}
\label{sec:real_axis}
\begin{figure*}
	\includegraphics[width=0.98\linewidth]{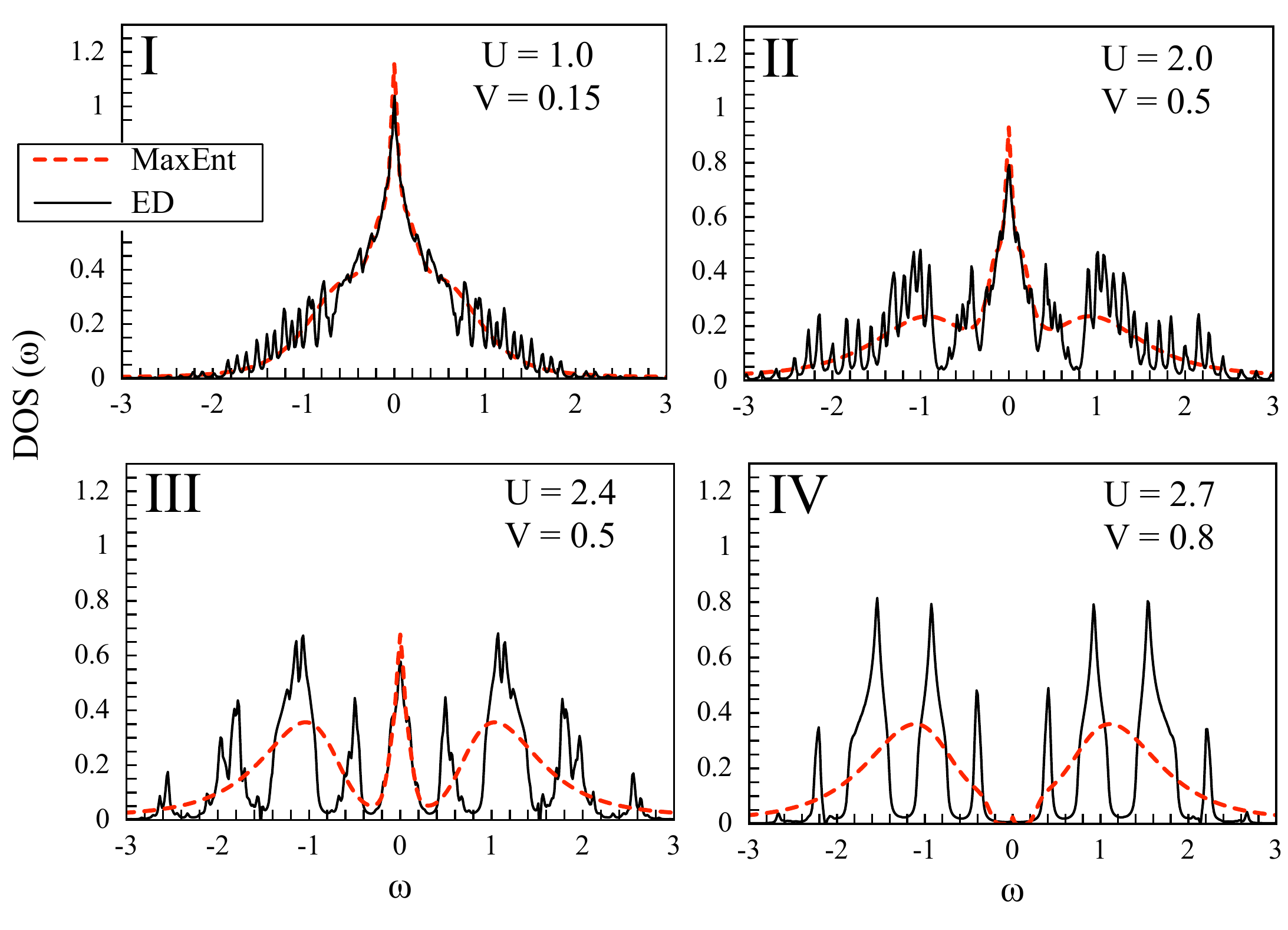}
    \caption{\label{fig:dos_qmc_ed}(Color online)
        Density of states $\text{DOS}(\omega)=-\text{Im}G_{loc}(\omega+i\frac{\pi}{2\beta})/\pi$ obtained in ED (bold line) and from the CTQMC results via a conventional maximum entropy method (dashed line, cf. text).
        The values of $U$ and $V$ correspond to the points I-IV in the phase diagram in Fig.~\ref{PDV}.
}
\end{figure*}
The hallmark of ED methods is their direct access to the real axis.
Here one has the choice to obtain the density of states $\text{DOS}(\omega)$ at real $\omega$ either from the local lattice Green's function, $G_{loc}(\omega+i\delta)$, or from the impurity, $G(\omega+i\delta)$.
These yield a different $\text{DOS}(\omega)$, since in ED the self-consistency $G_{loc}(i\omega)=G(i\omega)$ can only be achieved approximately, due to the discretization of the fermionic bath.
However, only the local lattice Green's function $G_{loc}(i\omega)=N^{-1}\sum_{\kv}\left[i\omega+\mu-t(\kv)-\Sigma(i\omega)\right]^{-1}$ has the non-interacting system as an exact limit, i.e., when the self-energy $\Sigma$ vanishes.
This is not the case for the impurity Green's function $G(i\omega)=\left[i\omega+\mu-\Delta(i\omega)-\Sigma(i\omega)\right]^{-1}$,
which suffers from artifacts due to the discretization of the hybridization function $\Delta$ even for vanishing interaction.

In Fig.~\ref{fig:dos_qmc_ed} we show the density of states $\text{DOS}(\omega)=-\text{Im}G_{loc}(\omega+i\frac{\pi}{2\beta})/\pi$ obtained
from ED at the points marked with I-IV in the phase diagram in Fig.~\ref{PDV}.
We note that the ED results of Fig.~\ref{fig:dos_qmc_ed} correspond to the setup of the benchmarks of Sec.~\ref{sec:benchmarks}, which were performed with $K=7$ fermionic and $P=3$ bosonic bath levels.
We compare the DOS obtained in ED to the one obtained from CTQMC by a conventional maximum entropy method.~\cite{PhysRevB.57.10287}
To this end, we choose the distance $i\delta=i\frac{\pi}{2\beta}$ to the real axis in ED, achieving a good match of the quasiparticle peak between both methods.
We stress that \textit{both} methods yield a DOS on the real axis that is only reliable close to the Fermi level, while the shape and number of the peaks away from the Fermi level is uncertain.
This is the case in ED due to artifacts stemming from the discretization of the fermionic bath, which are small only where the self-energy $\Sigma$ is small (cf. discussion above).
The conventional maximum entropy method, on the other hand, is not reliable in reproducing plasmon satellites in the DOS of the
Hubbard-Holstein model.~\cite{PhysRevB.85.035115,PhysRevB.90.195114}

\subsection{Bosonic bath and screening}
The question of screening of the local Hubbard repulsion $U$ by the nonlocal potential $V$ has been discussed extensively, see,
e.g., Refs.~\onlinecite{PhysRevLett.109.126408,PhysRevB.87.125149,PhysRevB.90.195114,PhysRevLett.111.036601}.
An analysis by means of a variational principle~\cite{PhysRevLett.111.036601} predicts in a simple approximation a linear screening mechanism, with $U^*=U-V$ as the screened interaction.
From this point of view the nonlocal interaction incentivizes the creation of a double occupancy,
and hence a gain $U$ in the potential energy, by a reduction $V$ of the potential energy on a neighboring
site.~\cite{PhysRevLett.111.036601}
A more detailed investigation found that the efficiency of screening depends on the physical regime with $U^*=U-\alpha V$.
The renormalization factor $\alpha$ goes to $1$ with increasing Hubbard repulsion $U$, i.e., according to the variational principle, a
linear screening proportional to $V$ is realized best in the Mott insulator.~\cite{PhysRevB.94.165141}
In EDMFT, on the other hand, $U(i\nu)=U+\Lambda(i\nu)$ is the screened interaction, where $\Lambda(i\nu)$, Eq.~\eqref{eq:lambda}, is the
dynamic interaction arising from the effective bosonic bath in the Anderson-Holstein impurity model, Eq.~\eqref{imp}.
The EDMFT approximation summarized in Sec.~\ref{sec:edmft} can be motivated from an exact limit of large dimensionality $D$,
where one sees that on the Bethe lattice with nearest neighbor interactions $v_0/\sqrt{2D}$ the effective dynamic interaction is given by
the impurity susceptibility, $\Lambda(i\nu)=v_0^2X(i\nu)$.~\cite{PhysRevLett.77.3391}
Screening is thus suppressed quadratically for small $v_0$ and the typical energy scale of the screening, the screening
frequency,~\cite{PhysRevLett.109.126408}
\begin{align}
   \Omega_s&=\frac{\int_0^\infty d\nu \text{Im}\Lambda(\nu+i\delta)/\nu}{\int_0^\infty d\nu\text{Im}\Lambda(\nu+i\delta)/\nu^2},
    \label{eq:omega_s}
\end{align}
is determined by the energy scale of the local charge excitations.
Due to the degradation of the quasiparticle peak in the MI regime, charge excitations correspond exclusively to transitions between the
bands above and below the Fermi level (i.e. Hubbard bands and plasmon satellites).
Charge excitations are thus shifted to high energies, the result is a large screening frequency of the order of the band gap.
This physical situation is similar in finite dimensions, where it was found that EDMFT predicts only weak screening effects in the
insulating phase.~\cite{PhysRevB.87.125149}
However, the application of strong-coupling CTQMC, which was used in most of our references, is difficult deep in the half-filled insulator (cf. discussion in Sec.~\ref{sec:benchmarks}).
Since ED can be applied as the impurity solver in this regime, we supplement the discussion in Ref.~\onlinecite{PhysRevB.87.125149}
and Ref.~\onlinecite{PhysRevB.90.195114} by comparing screening in the FL and in the MI.
\begin{figure}
	\includegraphics[width=0.98\linewidth]{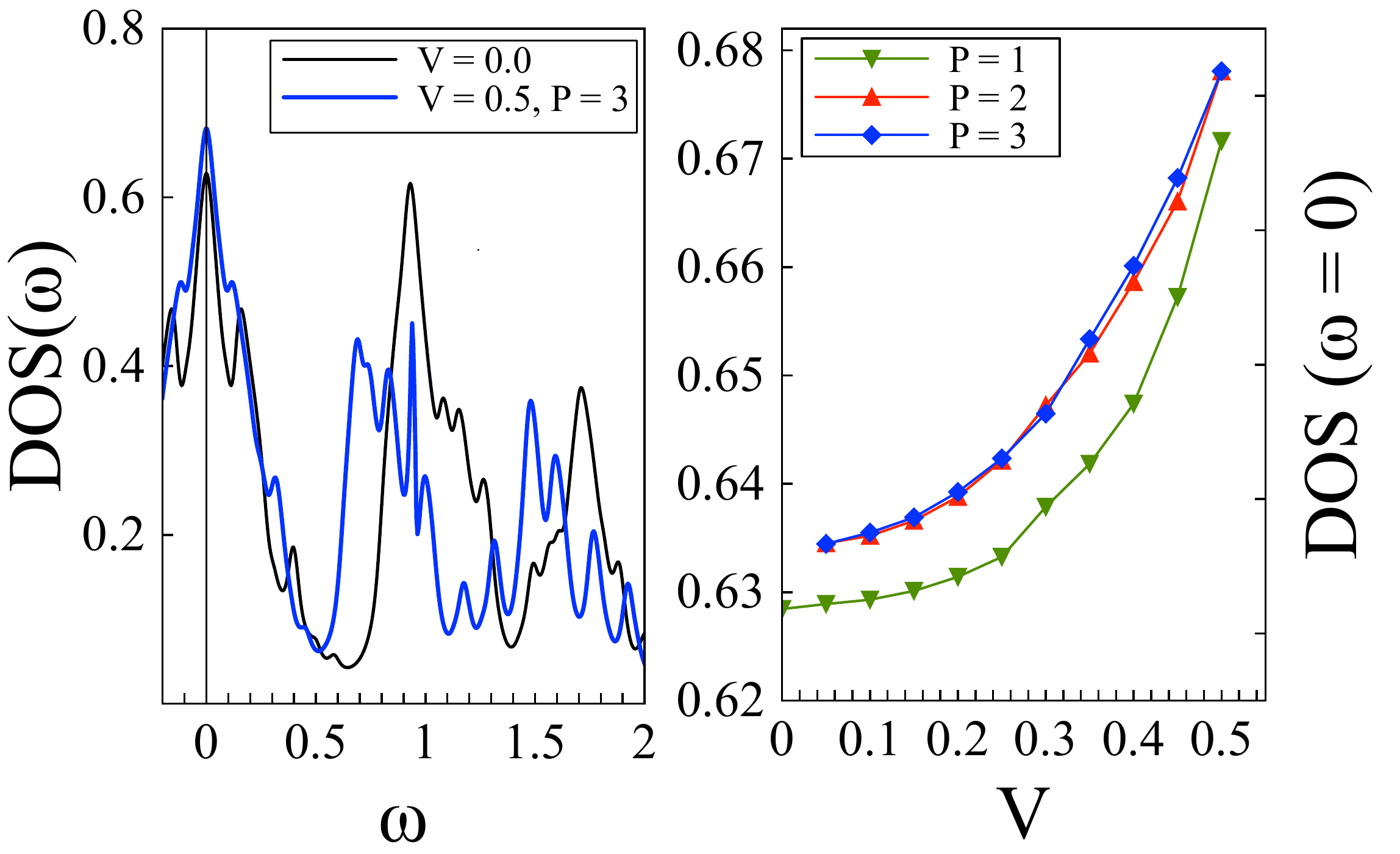}
    \caption{\label{fig:dos_metal}(Color online) Left: Spectral features of the DOS corresponding to the quasiparticle peak and the upper Hubbard band at $U=2$ with and without nonlocal interaction $V$.
        Right: DOS at the Fermi level for different values of $V$ (cf. blue diamonds in Fig.~\ref{PDV}). $P$ indicates how many energy levels were used for the discretization of the bosonic bath, the broadening of the DOS was chosen as $\delta=\pi/\beta$.
    }
\end{figure}

In Fig.~\ref{fig:dos_metal} we see the effect of $V$ on the DOS in the moderately correlated FL at $U=2$.
A series of calculations for $0\leq V\leq0.5$ in the right panel of Fig.~\ref{fig:dos_metal} shows a rise in the spectral weight at the Fermi level with $V$.
As can be seen in the left panel comparing $V=0$ and $V=0.5$, the nonlocal interaction screens the Hubbard repulsion and peaks corresponding to the upper Hubbard band are shifted towards the quasi-particle peak.
In order to assess the effect of the discretization of the bosonic bath, we performed the calculations with $P=1,2$ and $3$ bosonic bath levels,
keeping the number of fermionic bath levels fixed at $K=5$.
As we see from the right panel of Fig.~\ref{fig:dos_metal}, using $P=1$ or $2$ bosonic bath levels has a visible influence on the spectral weight at the Fermi level,
while there is a negligible difference between $P=2$ and $3$.

\begin{figure}
	\includegraphics[width=0.98\linewidth]{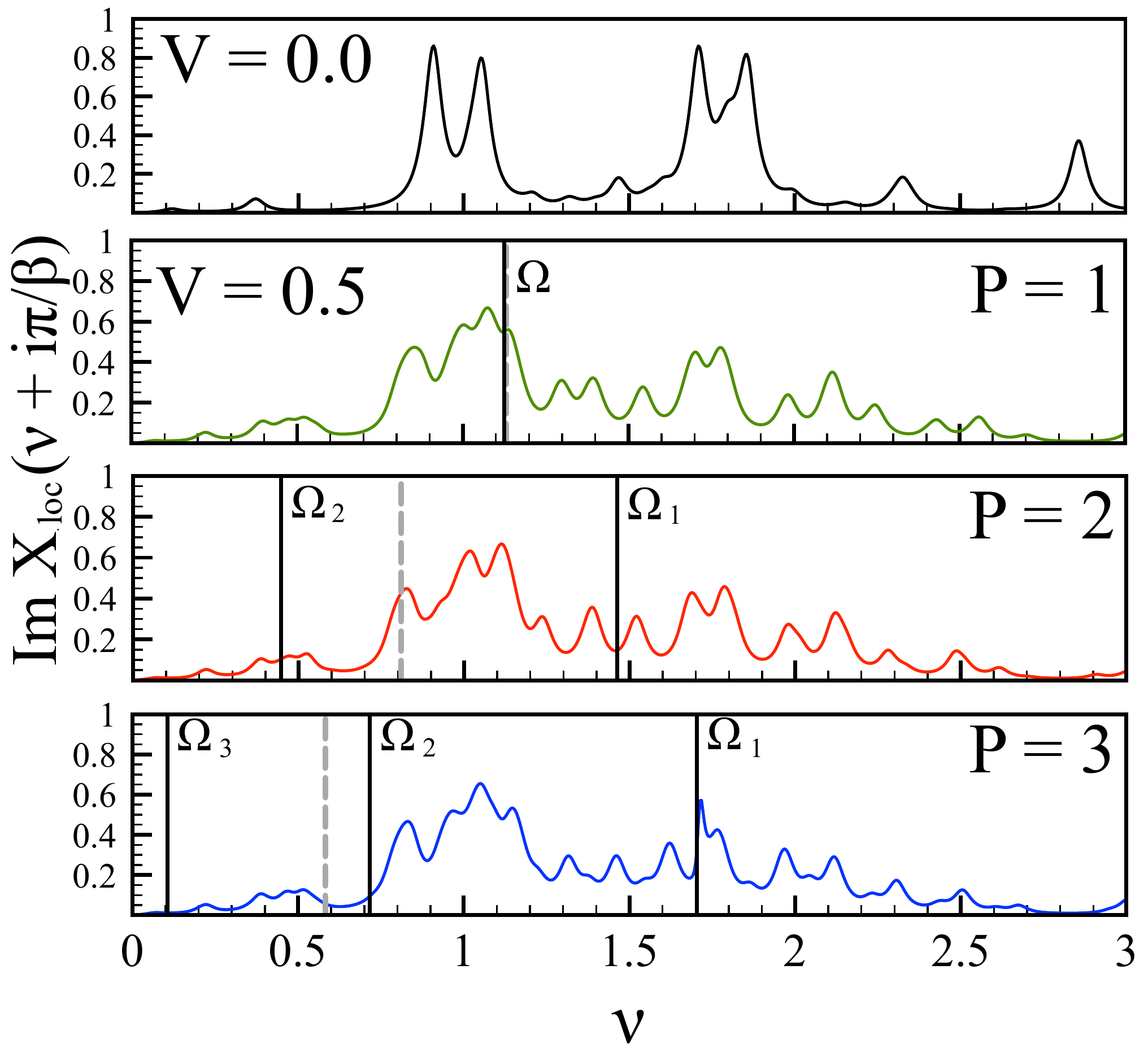}
    \caption{\label{fig:chi_metal}(Color online) Spectra of the local charge excitations for $U=2$ without nonlocal interaction (top panel) and at $V=0.5$ (lower panels).
        Bold vertical lines indicate the self-consistent modes $\Omega_p$ of the bosonic bath, they count to the number of bosonic bath levels $P$ used.
        Vertical dashed lines mark the screening frequency $\Omega_s$ from Eq.~\eqref{eq:omega_s}.
    }
\end{figure}

We see in Fig.~\ref{fig:chi_metal} the origin of the difference between $P=1$ and $P>1$, where we show the sprectrum of local charge
excitations on the real axis, $\text{Im}X_\text{loc}(\nu+i\delta)$. As discussed above, in EDMFT one can expect the energy scale of the
screening modes $\Omega_{p}$ that contribute to the dynamic interaction $\Lambda$ to be determined by the typical energies of the local
charge excitations. We have therefore drawn the $\Omega_{p}$ into their spectrum in Fig.~\ref{fig:chi_metal} for comparison. As becomes
clear from the second panel for $P=1$, a single bosonic bath energy $\Omega$ is insufficient to account for the broad spectrum of charge
excitations in the FL at $U=2, V=0.5$. Already at $P=2$ a much smaller $\Omega_2$ accounts for low energy charge excitations (cf. third
panel of Fig.~\ref{fig:chi_metal} and the second line of Table~\ref{table:P123}). This contribution to $\Lambda$ is important because low
energy modes are amplified in their contribution to screening effects on the DOS.~\cite{PhysRevB.90.195114} The low energy modes accounted
for when $P>1$ are therefore responsible for the difference in the spectral weight at the Fermi level in the right panel of
Fig.~\ref{fig:dos_metal} when going from $P=1$ to $P=2,3$ bosonic bath levels. Comparing the top panel of Fig.~\ref{fig:chi_metal} for $V=0$
with the bottom panels for $V=0.5$, we can also see that the low energy excitations are themselves enhanced by the nonlocal interaction $V$.
This corresponds to the shift of spectral weight closer to the Fermi level in the left panel of Fig.~\ref{fig:dos_metal}, due to the
screening of the Hubbard repulsion $U=2$ by $\Lambda(i\nu_0)\sim-0.3$, making the transition energies between the quasiparticle peak and the
upper Hubbard band smaller.
\begin{figure}
	\includegraphics[width=0.98\linewidth]{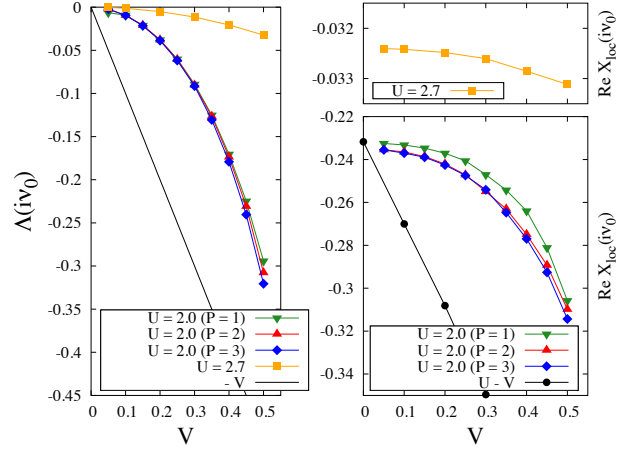}
    \caption{\label{fig:chi_lambda}(Color online) Left: Dynamic interaction $\Lambda$ at the lowest Matsubara frequency $i\nu_0=0$ as a function of $V$ in the FL (triangles and diamonds)
        and in the insulator (yellow squares, cf. yellow symbols in Fig.~\ref{PDV}). The line $-V$ indicates the static screening predicted
        by a simplistic evaluation of the variational principle presented in Ref.~\onlinecite{PhysRevLett.111.036601}.
        Top right: Local susceptibility $X_{\text{loc}}(i\nu_0)$ in the insulator.
        Bottom right: $X_{\text{loc}}(i\nu_0)$ in the FL, black symbols correspond to calculations at $U^*=U-V$.
        $P=1$ bosonic bath levels were used in the MI, in the FL we compare results for $P=1,2,3$.
    }
\end{figure}
Vertical dashed lines in Fig.~\ref{fig:chi_metal} indicate the screening frequency, Eq.~\eqref{eq:omega_s}, which for a finite number of
bosonic bath levels is given analytically as $\Omega_s=\sum_p(\mathcal{W}_p^2/\Omega_p)/\sum_p(\mathcal{W}_p^2/\Omega_p^2)$. We can see that
its value is drawn strongly to the lowest mode $\Omega_p$ and is not converged at $P=2$ bosonic bath levels.

In order to compare the effects of screening in the FL and MI we draw the dynamic interaction $\Lambda(i\nu)$ at $i\nu_0=0$ as a function of
$V$ in the left panel of Fig.~\ref{fig:chi_lambda}. In the FL for $U=2$ we find a similar behavior of $\Lambda(i\nu_0)$ as
in Ref.~\onlinecite{PhysRevB.90.195114}, showing that EDMFT does not behave like the variational estimate $U^*=U-V$. This discrepancy is
even more dramatic for screening in the insulator at $U=2.7$, where $\Lambda(i\nu_0)/(-V)\ll 1$. In contrast to the FL, where low energy charge
excitations serve as a mediator for significant screening effects of $V$ on the DOS, charge excitations in the MI are shifted to high
energies. Therefore, the bosonic bath energies $\Omega_p$ remain large in the MI for different numbers $P=1,2,3$ of bosonic bath levels (cf.
last line of Table~\ref{table:P123}). We can confirm from our calculations that for values of $U$ and $V$ in the MI, far enough from phase
boundaries, spectral features in the DOS within a broad region around the Fermi level are virtually immune to screening effects. We observe
in the top right panel of Fig.~\ref{fig:chi_lambda} that in such a case low energy charge excitations in turn receive almost no enhancement
by $V$. In the FL, on the other hand, low energy charge excitations are amplified by $V$, as seen in the bottom right panel of
Fig.~\ref{fig:chi_lambda}. Already at small $V$ at least one low energy mode $\Omega_p$ and thus $P>1$ is needed to account for these
excitations (cf. discussion on Fig.~\ref{fig:chi_metal} and the second line of Table~\ref{table:P123}).

\section{Conclusions}
\label{sec:conclusions}

We have presented a solver for the Holstein-Anderson impurity model that is based on the exact diagonalization (ED) technique
and applied it to solve the EDMFT equations for the extended Hubbard model on the square lattice with a nearest-neighbor interaction.
The discretization of a fermionic and of a bosonic hybridization function is necessary in ED. Benchmarks with a CTQMC solver in different
regimes show that in most cases two screening modes suffice for the discretization of the bosonic bath. This is consistent with prior
results that found two dominant screening modes in different regions of the EDMFT phase diagram.~\cite{PhysRevB.90.195114} An exception is
the boundary between the insulating and the charge-ordered phase, where more than two modes are needed due to the broad spectrum of charge
excitations near the transition. More significant than the discretization of the bosonic bath are unavoidable and familiar artifacts due to
the discretization of the fermionic bath, e.g., that either the Fermi liquid or the insulating phase are favored when using an odd or an
even number of fermionic bath sites, respectively. We nevertheless obtain phase boundaries that are in good agreement with prior
publications that used strong-coupling CTQMC as a solver for the Hubbard-Holstein model.

We have found that deep in the half-filled insulator ED can be well applied as a solver for the EDMFT equations, while it becomes difficult
to obtain converged solutions with a strong-coupling CTQMC solver. This has motivated us to apply ED in this regime, our results supplement
a comparison in Ref.~\onlinecite{PhysRevB.90.195114} of the screening effects predicted by EDMFT and by a simplistic evaluation of a
variational principle presented in Ref.~\onlinecite{PhysRevLett.111.036601}. At large enough distance from phase boundaries, EDMFT predicts
that in the Mott phase the extended Hubbard model is virtually immune to effects on the low energetic single-particle spectrum by a screening of the local by the
nonlocal interaction.

It seems likely to us that EDMFT underestimates screening in the half-filled insulator: The dynamically screened interaction in the local
impurity problem of EDMFT accounts merely for an average effect of the nonlocal potential exerted by charges on neighboring sites on the
impurity itself. Hence, when a double occupancy is created on the impurity, the local reference system of EDMFT accounts fully for the gain
in potential energy on the impurity due to a slightly screened local interaction, but is blind for a simultaneous reduction in potential
energy by the creation of a vacancy on a neighboring site. This mechanism, however, is the physical interpretation of the variational
principle and it leads in a first approximation to a screening which is linear in the nonlocal potential. EDMFT seems to ignore this effect
and predicts linear screening neither in the Fermi liquid nor in the Mott phase.

Evidence for a poor description of the insulator by EDMFT was given in a recent publication that stressed the importance of the Fock
exchange diagram on the phase boundary between the charger-order and the Mott phase.~\cite{PhysRevB.95.245130} In a diagrammatic expansion
around EDMFT it can be expected that, in order to achieve a different local screening mechanism, a self-consistent renormalization of the
dynamically screened local interaction is needed. Indeed, a linear screening mechanism was found in the Fermi liquid phase by means of a
self-consistent version of the dual boson approach.~\cite{PhysRevB.93.045107} A further investigation of screening in the insulator seems
necessary to us and the proposed ED solver may help entering this regime.

\section{Acknowledgments}
F.K. likes to thank E.G.C.P. van Loon for fruitful discussions.
The ED calculations were done using the open source exact-diagonalization library,~\cite{2017arXiv170105645I}
based on the core libraries~\cite{Gaenko2017235} of the open source ALPS package.~\cite{1742-5468-2011-05-P05001}
The strong-coupling CTQMC results were obtained from open-source CTQMC solver.~\cite{HAFERMANN20131280}
D.M. and V.V.M. were supported by a grant of the President of the Russian Federation MD-6458.2016.2. S.I. was supported by the Simons
collaboration on the many-electron problem and by Act 211 Government of the Russian Federation, contract No. 02.A03.21.0006.  F.K. and
A.I.L. are financially supported by the DFG-SFB668 program. This research uses computational resources of the HLRN-cluster under Project No.
hhp 00040 and resources of the National Energy Research Scientific Computing Center, a DOE Office of Science User. Facility supported by the
Office of Science of the U.S. Department of Energy under Contract No. DE-AC02-05CH11231.

\appendix

\section{Alternative way of calculating $\chi(\nu_0 = 0)$}
We propose a way to calculate a bosonic correlation function $\chi(i\nu)=\langle\hat{O}\hat{O}\rangle_\nu$, where $\hat{O}$ is a bosonic operator,
at $i\nu_0=0$ by means of ED, without evaluating the Lehmann expression, Eq.~\eqref{two-particleGF}, at this point.
From the Fourier transform of $\chi(i\nu)$, $\chi(\tau) = {\beta}^{-1} \sum_{\nu} \chi(i\nu) e^{-i\nu\tau}$, one obtains at $\tau=0$,
\begin{eqnarray}
    \chi(i\nu_0)=\beta\chi(\tau=0)-\sum_{\stackrel{n=-\infty}{n\neq0}}^\infty\chi(i\nu_n)\label{alternative_nu0}.
\end{eqnarray}
The equal-time expectation value $\chi(\tau=0)=\langle\hat{O}\hat{O}\rangle$ is easily determined in ED calculations, whereas the infinite sum over Matsubara frequencies on the right-hand-side converges relatively slowly.
However, a summation over the bosonic Matsubara frequencies corresponds to a contour integration $(2\pi i)^{-1}\oint\chi(z)f(z)dz, z\in\mathbb{C}$, where $f$ is the Bose distribution function $f(z)=[e^{\beta z}-1]^{-1}$.
The bosonic Matsubara energies $i\nu_n=2n\pi i/\beta$ are the poles of $f(z)$ with the residues $1/\beta$.
There exists a continued fraction representation $f_L(z)$ with a finite number of $2L+1$ poles $i\tilde{\nu}_l$ with residues $r_l/\beta$,
where $-L\leq l\leq L$.~\cite{PhysRevB.75.035123} This continued fraction $f_L(z)$ converges to $f(z)$ with increasing $L$, $f_{L\rightarrow\infty}=f$.
For $L<\infty$ the poles $i\tilde{\nu}_l$ are at different locations on the imaginary axis than the Matsubara energies $i\nu_n$, except for $i\tilde{\nu}_0=i\nu_0=0$,
where the residue of the continued fraction $f_L$ is $r_0/\beta=1/\beta$.
Therefore, using $f_L\approx f$ for finite $L$ in the contour integration, one writes the infinite sum in Eq.~\eqref{alternative_nu0} as
\begin{eqnarray}
\sum_{\stackrel{n=-\infty}{n\neq0}}^\infty\chi(i\nu_n)\approx\sum_{\stackrel{l=-L}{l\neq0}}^{L}r_l\chi(i\tilde{\nu}_l).
\end{eqnarray}
The error of this approximation approaches the machine error already for small $L$. 
In ED one may hence calculate $\chi$ at the alternative poles $i\tilde{\nu}_{l\neq0}$ to obtain $\chi(i\nu_0=i\tilde{\nu}_{0})$ from Eq.~\eqref{alternative_nu0}.
A separate implementation of the Lehmann expression, Eq.~\eqref{two-particleGF}, for the point $i\nu_0=0$ is then no longer necessary.
The alternative poles $i\tilde{\nu}_l$ and residues $r_l/\beta$ are provided online\protect{\footnote{A program to compute alternative poles and residues for the Fermi distribution function at $\beta=1$ is provided by the author of \cite{PhysRevB.75.035123} under the link
\href{http://www.openmx-square.org/zero_fermi/zero_fermi.c}{www.openmx-square.org/zero$\_$fermi/zero$\_$fermi.c}.
The provided program solves a generalized eigenproblem of the form $Ax = \lambda Bx$, where $B_{kl}=(2k-1)\delta_{kl}$.
The alternative poles and residues for the Bose distribution function are obtained for $B_{kl}=(2k+1)\delta_{kl}$ and may be scaled with $1/\beta$ to adjust the temperature.}}.

\bibliographystyle{apsrev4-1}
\bibliography{refs}

\end{document}